\documentclass[conference]{IEEEtran}
\renewcommand{\baselinestretch}{0.99}
\usepackage{ifpdf}
\usepackage{url}
\ifpdf
\else
\fi

\usepackage{cite}
\usepackage{balance}
\usepackage{bm,comment,color}

\ifCLASSINFOpdf
  \usepackage[pdftex]{graphicx}
  \graphicspath{{../pdf/}{../jpeg/}}
  \DeclareGraphicsExtensions{.pdf,.jpeg,.png}
\else
  \usepackage[dvips]{graphicx}
  \graphicspath{{../eps/}}
  \DeclareGraphicsExtensions{.eps}
\fi
\usepackage{amsmath}
\usepackage{algpseudocode}
\algtext*{EndWhile}
\algtext*{EndIf}
\algtext*{EndFor}
\usepackage{algorithm}
\usepackage{array}
\usepackage{amsfonts} 
\usepackage{amssymb}
\usepackage{esint} 
\usepackage{units}
\usepackage{multirow}
\usepackage{comment}

\usepackage{color}
\newcommand{\red}[1]{{\color{red}{#1}}} 
\newcommand{\blue}[1]{{\color{blue}{#1}}} 

\usepackage{pgfplots}
\pgfplotsset{compat=newest}
\usetikzlibrary{plotmarks}
\usetikzlibrary{arrows.meta}
\usepgfplotslibrary{patchplots}
\usepackage{grffile}
\newlength\fheight 
\newlength\fwidth 
\usepgfplotslibrary{fillbetween}

\usepackage{acro}

\DeclareAcronym{los}{
short=LOS,
long= line-of-sight,
}

\DeclareAcronym{pn}{
short=PN,
long= phase noise,
}
\DeclareAcronym{cfo}{
short=CFO,
long= carrier frequency offset,
}
\DeclareAcronym{pan}{
short=PAN,
long= power amplifier nonlinearity,
}
\DeclareAcronym{iqi}{
short=IQI,
long= in-phase and quadrature imbalance,
}
\DeclareAcronym{hwi}{
short=HWI,
long= hardware impairment,
}

\DeclareAcronym{mcrb}{
short=MCRB,
long= misspecified Cram\'er-Rao bound,
}
\DeclareAcronym{crb}{
short=CRB,
long= Cram\'er-Rao bound,
}
\DeclareAcronym{cp}{
short=CP,
long= cyclic prefix,
}
\DeclareAcronym{ue}{
short=UE,
long= user equipment,
}
\DeclareAcronym{bs}{
short=BS,
long= base station,
}
\DeclareAcronym{rfc}{
short=RFC,
long= radio-frequency chain,
}
\DeclareAcronym{aoa}{
short=AOA,
long= angle-of-arrival,
}
\DeclareAcronym{ofdm}{
short=OFDM,
long= orthogonal frequency-division multiplexing,
}
\DeclareAcronym{adc}{
short=ADC,
long= analog to digital converter,
}
\DeclareAcronym{dac}{
short=DAC,
long= digital to analog converter,
}
\DeclareAcronym{lo}{
short=LO,
long= local oscillator,
}
\DeclareAcronym{ula}{
short=ULA,
long= uniform linear array,
}
\DeclareAcronym{mc}{
short=MC,
long= mutual coupling,
}
\DeclareAcronym{lb}{
short=LB,
long= lower bound,
}
\DeclareAcronym{pdf}{
short=PDF,
long= probability density function,
}
\DeclareAcronym{sm}{
short=SM,
long= standard model,
}
\DeclareAcronym{tm}{
short=TM,
long= true model,
}

\DeclareAcronym{mm}{
short=MM,
long= mismatched model,
}

\DeclareAcronym{mmle}{
short=MMLE,
long= mismatched maximum likelihood estimation,
}
\DeclareAcronym{mle}{
short=MLE,
long= maximum likelihood estimation,
}
\usepackage{tabu,longtable}

\ifCLASSOPTIONcompsoc
 \usepackage[caption=false,font=normalsize,labelfont=sf,textfont=sf]{subfig}
\else
 \usepackage[caption=false,font=footnotesize]{subfig}
\fi
\usepackage{stfloats}
\usepackage[hidelinks]{hyperref}
\usepackage{xcolor}
\long\def\comment#1{}

\newfont{\bbb}{msbm10 scaled 700}

\newcommand{\hthickline}{\noalign{\hrule height 0.80pt}}

\newfont{\bb}{msbm10 scaled 1100}

\newcommand{\av}{{\bf a}}

\newcommand{\dv}{{\bf d}}

\newcommand{\fv}{{\bf f}}

\newcommand{\hv}{{\bf h}}

\newcommand{\nv}{{\bf n}}

\newcommand{\pv}{{\bf p}}

\newcommand{\sv}{{\bf s}}

\newcommand{\wv}{{\bf w}}

\newcommand{\xv}{{\bf x}}
\newcommand{\yv}{{\bf y}}

\newcommand{\Am}{{\bf A}}
\newcommand{\Bm}{{\bf B}}
\newcommand{\Cm}{{\bf C}}
\newcommand{\Dm}{{\bf D}}
\newcommand{\Em}{{\bf E}}
\newcommand{\Fm}{{\bf F}}

\newcommand{\Jm}{{\bf J}}

\newcommand{\Gc}{{\cal G}}

\newcommand{\etav}{\hbox{\boldmath$\eta$}}

\newcommand{\epsilonv}{\hbox{\boldmath$\epsilon$}}

\newcommand{\muv}{\hbox{\boldmath$\mu$}}

\newcommand{\thetav}{\hbox{$\boldsymbol\theta$}}

\newcommand{\Deltam}{\hbox{\boldmath$\Delta$}}

\newcommand{\Xim}{\hbox{\boldmath$\Xi$}}

\newcommand{\trace}{{\hbox{tr}}}

\renewcommand{\arg}{{\hbox{arg}}}

\begin{document}

\bstctlcite{IEEEexample:BSTcontrol}
\title{MCRB-based Performance Analysis of \\6G Localization under Hardware Impairments}
\author{Hui Chen\IEEEauthorrefmark{1}, Sina Rezaei Aghdam\IEEEauthorrefmark{1}, Musa Furkan Keskin\IEEEauthorrefmark{1}, Yibo Wu\IEEEauthorrefmark{1}\IEEEauthorrefmark{3}, Simon Lindberg\IEEEauthorrefmark{2},\\ Andreas Wolfgang\IEEEauthorrefmark{2}, Ulf Gustavsson\IEEEauthorrefmark{3}, Thomas Eriksson\IEEEauthorrefmark{1}, Henk Wymeersch\IEEEauthorrefmark{1}\\
\IEEEauthorrefmark{1}Chalmers University of Technology, Sweden, \IEEEauthorrefmark{2}Qamcom Research \& Technology, Sweden,\\
\IEEEauthorrefmark{3}Ericsson Research, Sweden\\
e-mail: hui.chen@chalmers.se
}

\maketitle

\begin{abstract}
Location information is expected to be the key to meeting the needs of communication and context-aware services in 6G systems. User localization is achieved based on delay and/or angle estimation using uplink or downlink pilot signals. However, hardware impairments (HWIs) distort the signals at both the transmitter and receiver sides and thus affect the localization performance. While this impact can be ignored at lower frequencies where HWIs are less severe,  modeling and analysis efforts are needed for 6G to evaluate the localization degradation due to HWIs.
In this work, we model various types of impairments and conduct a misspecified Cram\'er-Rao bound analysis to evaluate the HWI-induced performance loss. 
Simulation results with different types of HWIs show that each HWI leads to a different level of degradation in angle and delay estimation performance.

\end{abstract}


\IEEEpeerreviewmaketitle

\section{Introduction}
\label{sec:intro}
Localization will be an indispensable part of future communication systems, both to improve spatial efficiency and optimize resource allocation~\cite{di2014location}, but also to support high-accuracy context-aware applications such as the tactile internet, augmented reality, and smart cities~\cite{kwon2021joint, xiao2020overview}. By taking advantage of a large array dimension and wide bandwidth of high-frequency (e.g., mmWave and sub-THz) communication systems, high angular and delay resolution, and hence accurate position estimation can be achieved~\cite{chen2021tutorial}.
%
%
Most localization algorithms rely on accurate models of the received signals as a function of the channel parameters (angles, delays, Dopplers) of the propagation environment. The presence of \acp{hwi} such as \ac{pn}, \ac{cfo}, \ac{mc}, \ac{pan}, \ac{iqi}, distort the pilot signals. As a result, when algorithms derived from a mismatched model (i.e., without or with limited information about the \acp{hwi}), the localization performance is unavoidably affected.

There has been extensive research on the effect of \acp{hwi} in communication systems in terms of spectral efficiency analysis~\cite{kolawole2018impact}, beamforming optimization~\cite{shen2020beamforming}, and channel estimation~\cite{wu2019efficient, ginige2021untrained}. The research on localization and sensing considering HWIs is also catching up. The effect of PN on automotive radar~\cite{yildirim2019impact,gerstmair2019safe, siddiq2018phase},
mutual coupling on DOA estimation~\cite{ye2009doa}, IQI on mmWave localization~\cite{ghaseminajm2020localization}, and PAN on joint radar-communication (JRC) systems \cite{bozorgi2021rf} are discussed. 
In~\cite{ghaseminajm2021localization}, the impairments are jointly modeled using a HWI factor. Nevertheless, this factor is not able to capture the contribution of each individual HWI. Hence, critical questions, such as how much error will be caused by a mismatched model, and how much HWI we can tolerate for 5G/6G localization, remain unanswered.

In this work, we consider an \ac{ofdm}-based localization system with \acp{hwi}. 
The corresponding localization algorithm may or may not have knowledge about these \acp{hwi}, 
where in the latter case the localization algorithm operates under model mismatch, as it does not know the PAN and the residual impairments of PN, CFO, and MC.
We use the \ac{crb} to predict the performance in angle, delay, and position estimation under the different models, 
and employ the \ac{mcrb}~\cite{richmond2015parameter,roemer2020misspecified, fortunati2017performance} to quantify the estimation performance loss due to model mismatch. The results show that different types of impairments affect angle and delay estimation in different ways.

\begin{figure}
\centering
\includegraphics[width=0.9\linewidth]{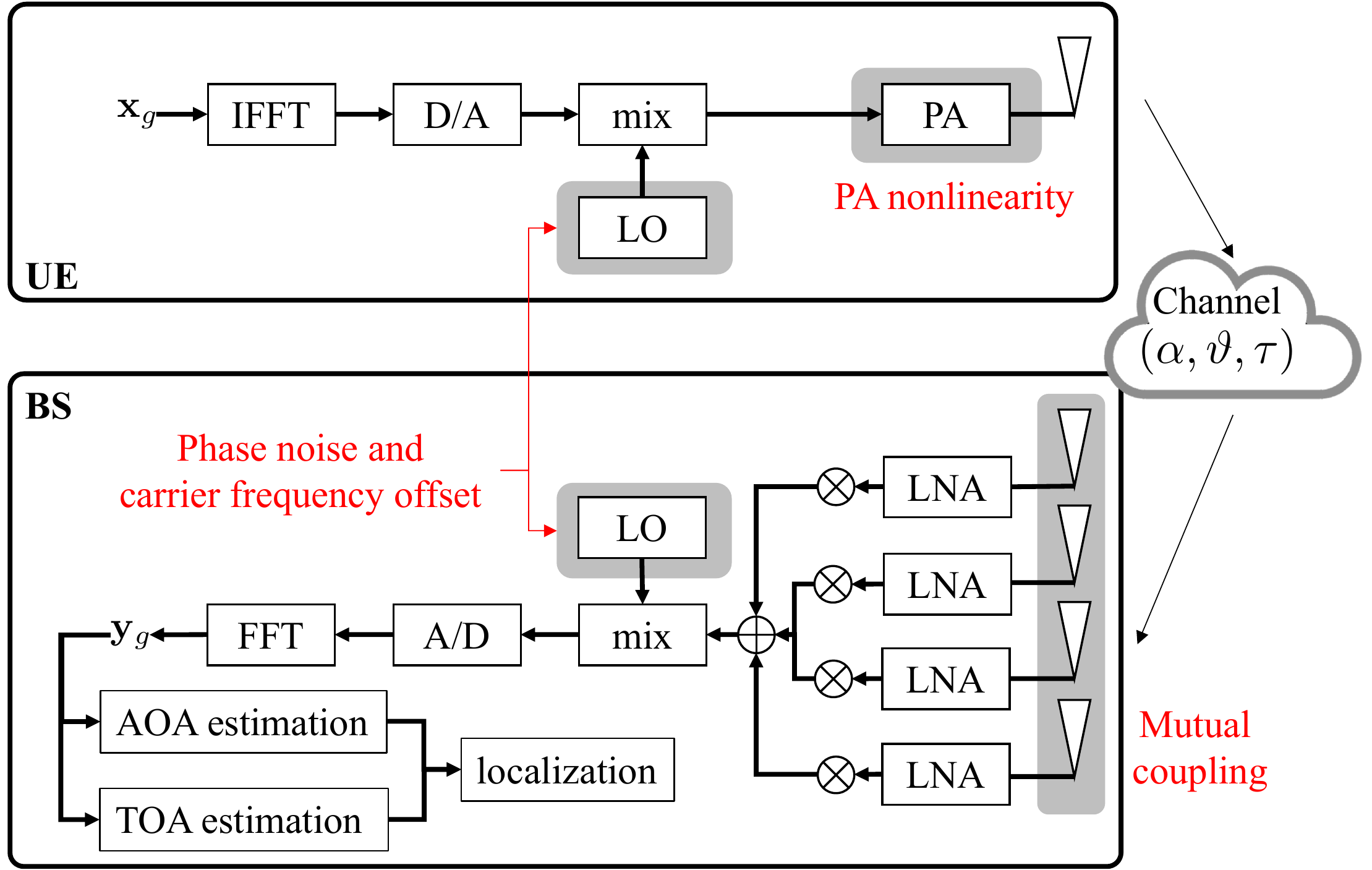}
\caption{Block diagram of considered hardware impairments (marked in gray) at transmitter and receiver. When the localization algorithm does not have perfect knowledge of the generative model, it operates under model mismatch.}
\label{fig:block_diagram}\vspace{-5mm}
\end{figure}


\section{System Model}
\label{sec-2-model}
In this section, we start with the \ac{hwi}-free model 
and then describe the \ac{hwi} model, as shown in Fig.~\ref{fig:block_diagram}. 
We consider a simplified uplink scenario with a \ac{los} channel between a \ac{bs} equipped with an $N_\text{}$-antenna \ac{ula} and a synchronized single-antenna \ac{ue}, both with a single \ac{rfc}. 
{The assumptions of single-antenna \ac{ue}, perfect synchronization, and pure \ac{los} may not be realistic in practice, but are an initial step to analyze and understand \acp{hwi}.}
We set the center of the BS array as the origin of the global coordinate system. The relation between the \ac{aoa} $\vartheta$, the delay $\tau$, and the UE position $\pv$ can be expressed as $\pv = {\tau}{c}\,[\cos(\vartheta), \sin(\vartheta)]^\top$, where $c$ is the speed of light.

\subsection{Hardware Impairment-free Model}

Considering the transmitted \ac{ofdm} symbol at $g$-th transmission ($1\le g \le \mathcal{G}$) and $k$-th subcarrier ($1\le k\le K$), $x_{g,k}$, its observation at the \ac{bs} 
can be formulated as
\begin{equation}
    y_{g,k} = \wv_{\text{}g}^\top \hv_k x_{g,k} + n_{g,k},
    \label{eq:ideal_far_field_model}
\end{equation}
where $\wv_{\text{}g} \in \mathbb{C}^{N}$ is the combiner at the BS for the \mbox{$g$-th} transmission; $\hv_k = \alpha D_k \av_\text{}(\vartheta)$ is the channel vector at the $k$th subcarrier with a complex gain (amplitude $\rho$ and phase $\xi$) as $\alpha = \rho e^{-j\xi} = {\lambda e^{-j\xi} }/{(4\pi c \tau)}$, an receiver steering vector $\av_\text{}(\vartheta) = [1, e^{j\pi\sin(\vartheta)}, \ldots, e^{j(N-1)\pi\sin(\vartheta)}]^\top$, 
and a delay component $D_k = e^{-j 2 \pi k \Delta_f \tau}$ ($\Delta_f$ is the subcarrier spacing). 
We assume $\hv_k$ remains the same during $\mathcal{G}$ transmissions (within the coherence time).
Finally, $n_{g,k}\in \mathcal{CN}(0, \sigma_n^2)$ is the noise following a complex normal distribution, with $\sigma_n^2=N_0 W$, where $N_0$ is the noise power spectral density (PSD) and $W=K\Delta_f$ is the total bandwidth. The average transmission power $P=\mathbb{E}\{ |x_{g,k}|^2\}/R$, where $R$ is the load impedance.  
By concatenating all the received symbols into a column, we obtain the received symbol block $\yv \in \mathbb{R}^{\mathcal{G}K}$ as $\yv = [\yv_1^\top, \ldots, \yv_g^\top, \ldots, \yv^\top_\mathcal{G}]^\top$, 
where $\yv_g=[y_{g, 1}, \ldots, y_{g, K}]^\top$ can be expressed as
\begin{align}
    \yv_g = \alpha\wv_{\text{}g}^\top  \av_\text{}(\vartheta) \dv(\tau) \odot \xv_{g} + \nv_{g} \label{eq:ideal_model_y_g}
\end{align}
in which $\dv(\tau)=[D_1,\ldots,D_K]^\top$, $\xv_g=[x_{g, 1}, \ldots, x_{g, K}]^\top$, $\nv_g=[n_{g, 1}, \ldots, n_{g, K}]^\top$ and $\odot$ denotes the Hadamard product. 


\subsection{Hardware Impairments}
The considered HWIs are highlighted in gray in Fig.~\ref{fig:block_diagram}~\cite{schenk2008rf}. We select residual PN, residual CFO, residual MC, and PAN. The \ac{iqi} and imperfections of \ac{adc}, \ac{dac}, low-noise amplifier (LNA) and mixer are left for future work.
By focusing on residual PN, CFO, and MC, our analysis can impose requirements on the corresponding PN, CFO, and MC estimation accuracy.

\subsubsection{Phase Noise and Carrier Frequency Offset} Imperfect \acp{lo} in the up-conversion and down-conversion processes add PN to the carrier wave phase. In addition, when the down-converting \ac{lo} in the receiver does not perfectly synchronize with the received signal’s carrier~\cite{mohammadian2021rf}, CFO occurs. Generally, both PN and CFO are tackled by the receiver~\cite{hajiabdolrahim2020extended}, so we only consider the residual PN and residual CFO at the receiver. With PN and CFO, the observation, $y_{g,k}$, is modified as~\cite{lin2006joint}
\begin{equation}
    y_{g,k} \to  \fv^\top_k \Em_g \Xim_g \Fm^{\mathsf{H}} \yv_g,
    \label{eq:PN_CFO}
\end{equation}
where $\yv_g$ is without PN or CFO (i.e., from \eqref{eq:ideal_far_field_model}), 
$\Fm = [\fv_1, \fv_2, \ldots, \fv_K]$ is the FFT matrix,
\begin{align}
    \Xim_g=\text{diag}([e^{j\omega_{g,1}}, \ldots, e^{j\omega_{g,K}}])
\end{align} is the residual\footnote{Since $\omega_{g,k}$ represents residual PN that remains after PN mitigation processing (e.g., \cite{salim2014channel,chung2021phase}), it is assumed to be independent across time.} phase noise matrix with $\omega_{g,k} \sim \mathcal{N}(0, \sigma_\text{PN}^2)$, and $\Em_g$ accounts for the CFO. 
In \eqref{eq:PN_CFO}, the vector $\yv_g$ is converted to the time domain by $\Fm^{\mathsf{H}} \yv_g$, where the successive phase noise samples, as well as the CFO are applied. Finally, $\fv^\top_k$ extracts the $k$-th subcarrier after applying an FFT  to $\Em_g \Xim_g \Fm^{\mathsf{H}} \yv_g$.
The CFO matrix $\Em_g$ considers both inter-OFDM symbol phase changes as well as inter-carrier interference~\cite{lin2006joint, roman2006blind}:
\begin{equation}
        \Em_g=e^{j\frac{2\pi \epsilon gK_{\text{tot}}}{{K}}}\text{diag}([1, e^{j\frac{2 \pi \epsilon}{K}}, \ldots, e^{j\frac{2 \pi (K-1) \epsilon}{K}}]),
\end{equation}
where $K_{\text{tot}}=K+K_\text{cp}$, in which 
$K_\text{cp}$ is the length of the cyclic prefix, and $\epsilon$ is the normalized residual CFO with $\epsilon \sim \mathcal{N}(0, \sigma_\text{CFO}^2)$.

\subsubsection{Mutual Coupling} \label{sec:coupling}
\Ac{mc} refers to the electromagnetic interaction between the antenna elements in an array~\cite{ye2009doa}. Similar to PN and CFO modeling, we consider residual \ac{mc}, which remains after a calibration procedure. 
{For a \ac{ula}, we introduce the banded \ac{mc} matrix $\Cm_\text{} \in \mathbb{C}^{N\times N}=\tilde\Cm_\text{}+\Deltam_\text{MC}$ at the Rx.
Here, $\tilde\Cm_\text{} = \text{Toeplitz}([1, c_1, \ldots, c_n, 0, \ldots, 0])$ is the MC matrix with $c_n$ as the MC coefficient, and $\Deltam_\text{MC}$ represents the uncalibrated MC matrix that is modeled as random (residual MC matrix) with each element $\Delta_{i,j} \sim \mathcal{N}(0, \sigma_\text{MC}^2))$.}
The MC leads to the substitution~\cite{ye2009doa}
\begin{equation}
    \hv_k\to \Cm_\text{}\hv_k.
    \label{eq:MC}
\end{equation}

\subsubsection{Power Amplifier Nonlinearity}
For the PA nonlinearity, we consider a $Q$-th order memoryless polynomial nonlinear model with a clipping point  $x_\text{clip} \in \mathbb{R}$ as~\cite{schenk2008rf}
\begin{equation}
{h}_{\text{PA}}(\check{{x}}) =
\begin{cases}
      \sum_{q = 0}^{Q} \beta_{q} \check{{x}} |\check{{x}}|^{q} & |\check x| \le x_{\text{clip}},\\
      \sum_{q = 0}^{Q} \beta_{q} \frac{\check{{x}}}{|\check{{x}}|} |{{x}_\text{clip}}|^{q} & |\check x| > x_{\text{clip}},
\end{cases}
\label{eq:PA}
\end{equation}
where $\check{x}$ denotes the transmitted signals in time domain and $\beta_0, \dots, \beta_{Q}$
are complex-valued parameters. Note that the PA affects the time domain signals and we assume no digital pre-distortion is implemented. We also use non-oversampled signals as the input of the PA for tractable localization performance analysis. 

\subsection{Hardware Impaired Model}
Considering the \acp{hwi} of PN, CFO, MC, and PA nonlinearity and substituting~\eqref{eq:PN_CFO}, \eqref{eq:MC}, and~\eqref{eq:PA} into~\eqref{eq:ideal_model_y_g}, the observation can be rewritten in the frequency domain as
\begin{align}
    {{{\mathbf{y}}}}_g 
     = & \alpha \mathbf{F}
     \mathbf{E}_{g}\mathbf{\Xi}_{\text{}g}
     \mathbf{F}^{\mathsf{H}}(\mathbf{w}_{\text{}g}^{\top}\mathbf{C}_{\text{}} \av_\text{}(\vartheta)) \dv(\tau) \label{eq:impaired_model} \\
    & \odot (\mathbf{F}^{\top} \mathbf{h}_{\text{PA}}(\mathbf{F}^{\mathsf{H}} \mathbf{x}_g))  + {\mathbf{n}_{g}}
    \notag \\ \nonumber
      = &  \alpha \, \bar \etav_g(\pv)+ {\mathbf{n}_{g}} = \bar \muv_g(\alpha, \pv) + {\mathbf{n}_{g}}, 
\end{align}
where $\mathbf{h}_{\text{PA}}(\cdot)$ overloads the notation for the PA, and operates point-wise on each of the elements in the time-domain sequence. The FFT and IFFT matrices switch between time and frequency domain representations in order to apply the PN, CFO and PA in the correct (time) domain, while providing a frequency domain representation. We use $\bar{\muv}_g(\alpha, \pv)$ to denote the noise-free observation. Note that the PA model in~\eqref{eq:PA} does not consider the out-of-band emissions, but only the in-band distortion. 

Finally, we consider a model without 
the PAN and the residual noise of PN, CFO, and MC: 
\begin{align}
    \yv_g & = \alpha\wv_{\text{}g}^\top  \tilde\Cm_\text{} \av_\text{}(\vartheta) \dv(\tau) \odot \xv_{g} + \nv_{g}
    \label{eq:mismatched_model}
    \\ \nonumber
    = &  \alpha \, \etav_g(\pv)+ {\mathbf{n}_{g}} = \muv_g(\alpha, \pv) + {\mathbf{n}_{g}},
\end{align}
where $\muv_g(\alpha, \pv)$ is the noise-free version of the observation.

\subsection{Summary of the Models}
{To summarize, we have defined three types of signal models as follows.
\begin{itemize}
    \item \textbf{M0}: The model defined in~\eqref{eq:ideal_far_field_model}  without considering the \ac{hwi}. 
    \item \textbf{M1}: The model that considers knowledge of the various \acp{hwi} defined in~\eqref{eq:impaired_model}.
    \item \textbf{M2}: With the practical assumption that not all the \acp{hwi} information is available, we use the model defined in~\eqref{eq:mismatched_model}.
\end{itemize}
}
In the rest of the work, M0 will not be discussed, and the models M1 and M2 will be used for CRB analysis, as well as for localization performance evaluation. 

\section{Localization Algorithm}
\label{sec:localization_algorithm}
The \ac{mle} is employed when the observation $\yv$ is generated from the same model used by the algorithm. On the other hand, the \ac{mmle} is used when the observation $\yv$ is generated from a different model than what is used by the algorithm. In the latter case, we will denote the generative model by \emph{\ac{tm}}, while the model used by the estimator is called the \emph{\ac{mm}}. 
\subsection{MLE} 
If $\yv \sim f_\text{TM}( \yv|\alpha, \pv)$, 
the \ac{mle} of the UE position and channel gain is 
\begin{align}
        [\hat \pv_\text{MLE} ,\hat \alpha_\text{MLE}]
        & = \arg \max_{\pv,\alpha} \ln f_\text{TM}( \yv|\alpha, \pv),
\end{align}
where $\ln f_\text{TM}( \yv|\alpha, \pv)$ is the log-likelihood of the \ac{tm}.
Since $\alpha$ appears linearly in the noise-free observation, we can use a plug-in estimate, and solve for the position by a coarse grid search to find an initial estimate $\pv_0$, followed by a backtracking line search~\cite{nocedal2006numerical}. For instance, when $\text{TM}=\textbf{M1}$, 
\begin{align}
       \hat \pv_\text{MLE}& = \arg \min_\pv \left \Vert  \yv - \frac{\bar\etav(\pv)^{\mathsf{H}}  \yv}{\Vert \bar\etav(\pv)\Vert^2}\bar\etav(\pv) \right \Vert^2.
        \label{eq:MLE_perfect_HWI}
\end{align}


\subsection{MMLE} 
If $\yv \sim f_\text{TM}( \yv|\alpha, \pv)$, but the estimator uses $f_\text{MM}( \yv|\alpha, \pv) \neq f_\text{TM}( \yv|\alpha, \pv)$, the \ac{mmle} is given by
\begin{align}
        [\hat \pv_\text{MMLE} ,\hat \alpha_\text{MMLE}]
        & = \arg \max_{\pv,\alpha} \ln f_\text{MM}( \yv|\alpha, \pv),
\end{align}
For instance, when $\text{MM}=\textbf{M2}$ and  $\text{TM}\neq \textbf{M2}$,
\begin{align}
        \hat \pv_\text{MMLE}& = \arg \min_\pv \left \Vert  \yv - \frac{\etav(\pv)^{\mathsf{H}}  \yv}{\Vert \etav(\pv)\Vert^2}\etav(\pv) \right \Vert^2.
        \label{eq:MLE_mismatched}
\end{align}

\section{Lower Bounds Analysis}
\label{sec:lower_bounds_analysis}

In the next, we derive the CRB for M2\footnote{The CRB of M1 can be obtained similarly, which will not be detailed in this work.}, as well as the \ac{mcrb} for the mismatched estimator in~\eqref{eq:MLE_mismatched} with $\text{TM}= \text{M1}$ and $\text{MM}=\text{M2}$.

\subsection{CRB Analysis}
We define a channel parameter vector as ${\boldsymbol\theta} = [\vartheta, \tau, \rho, \xi]^\top$ and a state vector $\sv = [p_x, p_y, \rho, \xi]^\top$.
Given the signal model in~\eqref{eq:mismatched_model} and $\yv \sim f_
    {\text{\ac{mm}}}( \yv|\alpha, \pv)$, the CRB of the \ac{mm} can be obtained as~\cite{elzanaty2021reconfigurable}
\begin{equation}
    \mathrm{CRB} \triangleq \left[\mathbf{I}(\sv)\right]^{-1} = \left[\Jm_\mathrm{S}^\top \mathbf{I}({\boldsymbol\theta}) \Jm_\mathrm{S}\right]^{-1},
    \label{eq:CRB_from_FIM}
\end{equation}
where
\begin{align}
    \mathbf{I}({\boldsymbol\theta}) & 
    = \frac{2}{\sigma_n^2}\sum^{\Gc}_{g=1} \sum^K_{k=1}\mathrm{Re}\left\{
    \left(\frac{\partial\mu_{g,k}}{\partial{\boldsymbol\theta}}\right)^{\mathsf{H}} 
    \left(\frac{\partial\mu_{g,k}}{\partial{\boldsymbol\theta}}\right)\right\},
    \\
    \Jm_\mathrm{S} & \triangleq \frac{\partial {\boldsymbol\theta}}{\partial \sv} = 
    \begin{bmatrix}
        \frac{\partial \vartheta}{\partial p_x} & \frac{\partial \tau}{\partial p_x} & 0 & 0\\
        \frac{\partial \vartheta}{\partial p_y} & \frac{\partial \tau}{\partial p_y} & 0 & 0\\ 
        0 & 0 & 1 & 0\\ 
        0 & 0 & 0 & 1
    \end{bmatrix}.
    \label{eq:FIM_measurement}
\end{align}
Here, $\mathbf{I}({\boldsymbol\theta})$, $\mathbf{I}(\sv)$ are the Fisher information matrices (FIMs) of the channel parameter vector and UE state vector, $\mathrm{Re}\{\cdot\}$ is getting the real part of a complex number, and $\Jm_\mathrm{S}$ is the Jacobian matrix using a denominator-layout notation with ${\partial \vartheta}/{\partial \pv}={1}/{(c \tau)}[-\sin(\vartheta)\,\cos(\vartheta)]^\top$ and ${\partial \tau}/{\partial \pv}={\pv}/{(c \tau)}$.
Based on the FIM, we further define the angle error bound (AEB), delay error bound (DEB) and position error bound (PEB) as
\begin{align}
\mathrm{AEB} & = \sqrt{([\mathbf{I}({\boldsymbol\theta})^{-1}]_{1, 1})}
\label{eq:AEB},\\
\mathrm{DEB} & = \sqrt{([\mathbf{I}({\boldsymbol\theta})^{-1}]_{2, 2})}
\label{eq:DEB},\\
\mathrm{PEB} & = \sqrt{\trace([\mathrm{CRB}]_{1:2, 1:2})},
\label{eq:PEB}
\end{align}
where $\trace(\cdot)$ returns the trace of a matrix, and $[\cdot]_{i,j}$ is getting the element in the $i$th row, $j$th column of a matrix. The bounds from~\eqref{eq:AEB}--\eqref{eq:PEB} will be used to evaluate the localization performance.

\subsection{Misspecified CRB}
The model is said to be mismatched or misspecified when $\yv \sim f_
    {\text{TM}}( \yv|\alpha, \pv)$, while the estimation is based on 
     the assumption that $\yv \sim f_
    {\text{MM}}( \yv|\alpha, \pv)$), where $f_
    {\text{TM}}( \yv|\alpha, \pv)\neq f_
    {\text{MM}}( \yv|\alpha, \pv)$. 
Due to the one-to-one mapping between $\mathbf{s}$ and $\boldsymbol \theta$, we can also write $f_{\text{TM}}( \yv|\boldsymbol \theta)$ and $f_
    {\text{MM}}( \yv|\boldsymbol \theta)$.
The \ac{lb} of using a mismatched estimator can be obtained as~\cite{fortunati2017performance}
\begin{align}
    \text{LB}(\bar {\boldsymbol\theta}, {\boldsymbol\theta}_0) 
    & = \underbrace{\Am_{{\boldsymbol\theta}_0}^{-1}\Bm_{{\boldsymbol\theta}_0}\Am_{{\boldsymbol\theta}_0}^{-1}}_{=\text{MCRB}({\boldsymbol\theta}_0)} + \underbrace{(\bar{\boldsymbol\theta} - {\boldsymbol\theta}_0)(\bar{\boldsymbol\theta} - {\boldsymbol\theta}_0)^\top}_{=\text{Bias}({\boldsymbol\theta}_0)}, 
\end{align}
where $\bar{\thetav}$ is the true channel parameter vector, ${\boldsymbol\theta}_0$ is the pseudo-true parameter vector (which will be introduced soon), and $\Am_{{\boldsymbol\theta}_0}, \Bm_{{\boldsymbol\theta}_0}$ are two possible generalizations of the FIMs.
The LB is a bound in the sense that \blue{(typo `\red{$(\text{LB}(\bar {\boldsymbol\theta}, {\boldsymbol\theta}_0))^{-1}$}' is corrected as `\red{$\text{LB}(\bar {\boldsymbol\theta}, {\boldsymbol\theta}_0)$}')}
\begin{align}
    \mathbb{E} \{ (\hat{\boldsymbol{\theta}}_{\text{MMLE}} -\bar{\thetav})(\hat{\boldsymbol{\theta}}_{\text{MMLE}} -\bar{\thetav})^\top \} \succeq \text{LB}(\bar {\boldsymbol\theta}, {\boldsymbol\theta}_0),
\end{align}
where the expectation is with respect to $f_
    {\text{TM}}( \yv|\boldsymbol \theta )$.  What remains is the formal definition and computation of the pseudo-true parameter $\boldsymbol{\theta}_0$ and  $\Am_{{\boldsymbol\theta}_0}, \Bm_{{\boldsymbol\theta}_0}$.

\subsubsection{Pseudo-true Parameter}
{Assume the \ac{pdf} of the \ac{tm}, where the observation data come from, is $f_\text{TM}(\yv|\bar {\boldsymbol\theta})$, where $\yv$ is the received signals and $\bar {\boldsymbol\theta} \in \mathbb{R}^4$ (4 unknowns for this 2D case) is the vector containing all the channel parameters. Similarly, the \ac{pdf} of the \ac{mm} for the received signal $\yv$ can be noted as $f_\text{MM}(\yv, {\boldsymbol\theta})$.}
The pseudo-true parameter vector is defined as the point that minimizes the Kullback-Leibler divergence between $f_\text{TM}(\yv|\bar {\boldsymbol\theta})$ and $f_\text{MM}(\yv| {\boldsymbol\theta})$ as
\begin{align}
    {\boldsymbol\theta}_0 = \arg \min_{\boldsymbol\theta} D_\text{KL}(f_\text{TM}(\yv|\bar {\boldsymbol\theta})\Vert f_\text{MM}(\yv| {\boldsymbol\theta})).
\end{align}
We define $\epsilonv(\thetav) \triangleq \bar\muv(\bar {\boldsymbol\theta}) - \muv({\boldsymbol\theta})$, and the pseudo-true parameter can be obtained as (details can be found in  Appendix~\ref{appendix_B})
\begin{equation}
    {\boldsymbol\theta}_0 
    = \arg \min_{{\boldsymbol\theta}} \Vert \epsilonv(\thetav) \Vert^2 
    = \arg \min_{{\boldsymbol\theta}} \Vert \bar\muv(\bar {\boldsymbol\theta}) - \muv({\boldsymbol\theta})\Vert^2  
    \label{eq:pseudotrue_final}.
\end{equation}
Hence, ${\boldsymbol\theta}_0$ can be found by solving \eqref{eq:MLE_mismatched} with the observation $\yv = \bar\muv(\bar {\boldsymbol\theta})$, which can be accomplished using the same algorithm in Sec.~\ref{sec:localization_algorithm},
initialized with the true value $\bar {\boldsymbol\theta}$.
\subsubsection{MCRB Component Matrices}
The matrices $\Am_{{\boldsymbol\theta}_0}$ and $\Bm_{{\boldsymbol\theta}_0}$ can be obtained based on the pseudo-true parameter vector ${\boldsymbol\theta}_0$ as
\begin{align}
    &[\Am_{{\boldsymbol\theta}_0}]_{i,j}
        = \left. 
            \int \frac{\partial^2 \text{ln}f_\text{MM}(\yv|{\boldsymbol\theta})}{\partial{\theta}_i \partial{\theta}_j} f_\text{TM}(\yv|\bar {\boldsymbol\theta})\text{d}\yv
        \right|_{{\boldsymbol\theta} = {\boldsymbol\theta}_0} \notag \\
        &= \left.
            -\frac{1}{\sigma_n^2}\sum_{g,k}\int \frac{\partial^2|y_{g,k}-\mu_{g,k}(\boldsymbol\theta)|^2}{\partial \theta_i \partial \theta_j} f_\text{TM}(y_{g,k}|\bar\thetav)\text{d}y_{g,k} 
        \right|_{{\boldsymbol\theta} = {\boldsymbol\theta}_0}\notag 
        \\
        &=  \left. \frac{2}{\sigma_n^2}\text{Re}\left[\frac{\partial^2\muv({\boldsymbol\theta})}{\partial \theta_i \partial \theta_j}\epsilonv({\boldsymbol\theta}) -  \frac{\partial\muv({\boldsymbol\theta})}{\partial \theta_j}
        \left(\frac{\partial\muv({\boldsymbol\theta})}{\partial \theta_i} \right)^{\mathsf{H}}\right]
        \right|_{{\boldsymbol\theta} = {\boldsymbol\theta}_0}\label{eq:matrix_A}
\end{align}
and
\begin{align}
        &[\Bm_{{\boldsymbol\theta}_0} ]_{i,j} 
        = \left. \int 
        \frac{\partial \text{ln}f_\text{MM}(\yv|{\boldsymbol\theta})}{\partial{\theta}_i} 
        \frac{\partial \text{ln}f_\text{MM}(\yv|{\boldsymbol\theta})}{\partial{\theta}_j}
        f_\text{TM}(\yv|\bar {\boldsymbol\theta})\text{d}\yv
        \right|_{{\boldsymbol\theta} = {\boldsymbol\theta}_0} \notag \\
        & = 
        \frac{1}{\sigma_n^4}\int 
        \sum_{g,k}\sum_{g', k'}
        \frac{\partial |y_{g,k}-\mu_{g,k}(\thetav)|^2}{\partial{\theta}_i} 
        \notag  \\
        & \times 
        \frac{\partial |y_{g',k'}-\mu_{g',k'}(\thetav)|^2}{\partial{\theta}_j}
        f_\text{TM}(\yv|\bar {\boldsymbol\theta})\text{d} \yv
       \notag  \\
        & =  
        \frac{4}{\sigma_n^4}
        \text{Re} \left[
        \frac{\partial^2\muv({\boldsymbol\theta})}{\partial \theta_i}\epsilonv({\boldsymbol\theta})
        \right]
        \text{Re} \left[
        \frac{\partial^2\muv({\boldsymbol\theta})}{\partial \theta_j}\epsilonv({\boldsymbol\theta})
        \right] \notag  \\
        & \left. + 
        \frac{2}{\sigma_n^2}\text{Re}\left[  \frac{\partial\muv({\boldsymbol\theta})}{\partial \theta_j}
        \left(\frac{\partial\muv({\boldsymbol\theta})}{\partial \theta_i} \right)^{\mathsf{H}}
        \right]
        \right|_{{\boldsymbol\theta} = {\boldsymbol\theta}_0}.
    \label{eq:matrix_B}
\end{align}
where the calculation of each element inside the matrices  $\Am_{{\boldsymbol\theta}_0}$ and $\Bm_{{\boldsymbol\theta}_0}$ can be found in Appendix~\ref{appendix_C}.

\section{Numerical Results}
\label{sec-4-simulation}
\subsection{Default Parameters}
We consider a 2D uplink scenario with a single-antenna UE and a BS with a 20-element ULA. The pilot signal $x_{g,k}$ is chosen with a constant average energy $|x_{g,k}|^2 = {PR}$ and random phase.
The simulation parameters\footnote{The PA parameters are estimated from the measurements of the RF WebLab, which can be remotely accessed at \url{www.dpdcompetition.com}.} can be found in  Table~\ref{table:Simulation_parameters}.


\begin{table}[ht]
\scriptsize
\centering
\caption{Default Simulation Parameters}
\renewcommand{\arraystretch}{1.25}
\begin{tabular} {c | c | c }
    \hthickline
    \textbf{Parameters} & \textbf{True Model} & \textbf{Mismatched Model}\\
    \hline
    BS Antennas & \multicolumn{2}{c}{$N = 10$} \\
    \hline
    UE Antennas & \multicolumn{2}{c}{$1$}   \\
    \hline
    RFC at BS/UE & \multicolumn{2}{c}{$  1$}  \\
    \hline
    Carrier Frequency & \multicolumn{2}{c}{$f_c = \unit[140]{GHz}$}  \\
    \hline
    Bandwidth & \multicolumn{2}{c}{$W = \unit[1]{GHz}$}  \\
    \hline
    Transmissions & \multicolumn{2}{c}{$G = 10$}  \\
    \hline
    Subcarriers & \multicolumn{2}{c}{$K = 100$} \\
    \hline
    Length of the CP & \multicolumn{2}{c}{$K_\text{cp} = 7$} \\
    \hline
    Load Impedance & \multicolumn{2}{c}{$R = \unit[50]{\Omega}$} \\
    \hline
    Noise PSD & \multicolumn{2}{c}{$N_0 = \unit[-173.855]{dBm/Hz}$} \\
    \hline
    Noise Figure & \multicolumn{2}{c}{$\unit[10]{dBm}$} \\
    \hthickline
    PN (residue) & $\;\;\;\;\;\;\; \sigma_\text{PN} = 10^\circ\;\;\;\;\;\;\;$ & $\sigma_\text{PN} = 0^\circ$ \\
    \hline
    CFO (residue) & $\sigma_\text{CFO} = 0.01$ (\unit[0.71]{ppm}) & $\sigma_\text{CFO} = 0$ \\
    \hline
    MC (matrix) & \multicolumn{2}{c}{$c_1=$ $0.6$+$0.5j$, $c_2 =$ $0.4054$-$0.128j$}  
    \\
    \hline
    MC (residue) & $\sigma_\text{MC} = 0.02$ & $\sigma_\text{MC} = 0$ 
    \\
    \hline
    \multirow{2}{*}{PA Parameters}& \multicolumn{2}{c}{{$\beta_0=$ $0.9798$+$0.0286j$, $\beta_1=$ $0.0122$-$0.0043j$}}\\
    \multirow{2}{*}{}& \multicolumn{2}{c}{{$\beta_2=$ $-0.0007$+$0.0001j$}}\\
    \hline
    PA Clipping Voltage & \multicolumn{2}{c}{$x_\text{clip}=25V$}
    \\
    \hthickline
\end{tabular}
\renewcommand{\arraystretch}{1}
\label{table:Simulation_parameters}
\end{table}
\begin{figure}
\centering
%
%
\begin{tikzpicture}[scale=1\columnwidth/10cm,font=\footnotesize]
\begin{axis}[%
width=8cm,
height=4cm,
scale only axis,
xmin=0,
xmax=40,
xlabel style={font=\color{white!15!black}},
xlabel={$P$ [dBm]},
ymode=log,
ymin=0.00316227766016838,
ymax=2,
yminorticks=true,
ylabel style={font=\color{white!15!black}},
ylabel={Position RMSE [m]},
axis background/.style={fill=white},
xmajorgrids,
ymajorgrids,
yminorgrids,
legend columns=2, 
legend style={font=\footnotesize, at={(0.43,0.63)}, anchor=south west, legend cell align=left, align=left, draw=white!15!black}
]
\addplot [color=blue, dashed, line width=1.0pt, mark size=3.0pt, mark=o, mark options={solid, blue}]
  table[row sep=crcr]{%
-10	0.959148042199192\\
-6.875	0.964948932445966\\
-3.75	0.881353862155458\\
-0.625	0.786210788643783\\
2.5	0.572260854791976\\
5.625	0.371967797130524\\
8.75	0.247590398280124\\
11.875	0.127136386845837\\
15	0.0755959321330375\\
18.125	0.0579322031446538\\
21.25	0.0375750815129089\\
24.375	0.0304735334541091\\
27.5	0.0255443089738309\\
30.625	0.0215459163007383\\
33.75	0.0228204632380652\\
36.875	0.0219175575401332\\
40	0.0211201780327091\\
};
\addlegendentry{MMLE}

\addplot [color=red, dashed, line width=1.0pt, mark size=2.0pt, mark=square, mark options={solid, red}]
  table[row sep=crcr]{%
-10	0.961438495286098\\
-6.875	0.943381573588253\\
-3.75	0.861760951414449\\
-0.625	0.807704345130488\\
2.5	0.548036635875721\\
5.625	0.369303851483909\\
8.75	0.216217293896589\\
11.875	0.113413051070814\\
15	0.0728050935654174\\
18.125	0.053585989077988\\
21.25	0.0335213136968184\\
24.375	0.0253378278975343\\
27.5	0.0199703388909252\\
30.625	0.0169823178181243\\
33.75	0.0153824408330664\\
36.875	0.0144174094658256\\
40	0.0138340164290786\\
};
\addlegendentry{MLE-M1}

\addplot [color=black, line width=1.0pt, mark size=3.0pt, mark=+, mark options={solid, black}]
  table[row sep=crcr]{%
-10	1.41817160853921\\
-6.875	0.988140088651072\\
-3.75	0.688131327685669\\
-0.625	0.478841238190079\\
2.5	0.332913947299235\\
5.625	0.231250581467529\\
8.75	0.160556089029868\\
11.875	0.111576818096214\\
15	0.0778654059983426\\
18.125	0.0550041345831136\\
21.25	0.0399352328883187\\
24.375	0.0306347664091824\\
27.5	0.0253228611856299\\
30.625	0.0229311617916355\\
33.75	0.0221183367857013\\
36.875	0.0220324413575333\\
40	0.0222066223803488\\
};
\addlegendentry{LB}

\addplot [color=blue, line width=1.0pt]
  table[row sep=crcr]{%
-10	1.32385379090722\\
-6.875	0.923825665177739\\
-3.75	0.644673804239536\\
-0.625	0.449873097856311\\
2.5	0.313935206989789\\
5.625	0.219073589101788\\
8.75	0.152876250810251\\
11.875	0.106681723514101\\
15	0.0744457695137035\\
18.125	0.0519505348800905\\
21.25	0.0362526721391565\\
24.375	0.0252982234016005\\
27.5	0.0176538740322545\\
30.625	0.0123194132409705\\
33.75	0.00859686334707685\\
36.875	0.00599915417745099\\
40	0.00418639326831513\\
};
\addlegendentry{CRB-M2}

\addplot [color=red, line width=1.0pt]
  table[row sep=crcr]{%
-10	1.32781537507951\\
-6.875	0.925146231690372\\
-3.75	0.644188834103822\\
-0.625	0.448181308905151\\
2.5	0.311476377086928\\
5.625	0.21618510606921\\
8.75	0.149835409224311\\
11.875	0.103738748725679\\
15	0.0718513476845367\\
18.125	0.04998176890098\\
21.25	0.0352440720853085\\
24.375	0.025670044779373\\
27.5	0.0196651680848688\\
30.625	0.0165198915715537\\
33.75	0.0149370081919938\\
36.875	0.0141969776902201\\
40	0.0138559453558655\\
};
\addlegendentry{CRB-M1}

\end{axis}

\begin{axis}[%
width=1.25\fwidth,
height=1.25\fheight,
at={(-0.163\fwidth,-0.156\fheight)},
scale only axis,
xmin=0,
xmax=1,
ymin=0,
ymax=1,
axis line style={draw=none},
ticks=none,
axis x line*=bottom,
axis y line*=left
]
\end{axis}
\end{tikzpicture}%
\vspace*{-1cm}
\caption{Comparison between estimator results (MMLE and MLE-M1) and different lower bounds. Due to the HWIs, the performance saturates when the transmission power exceeds~\unit[30]{dBm}. However, with the knowledge of the impairments, the bound (red curve with square markers) could be lower than the LB (blue curve with circle markers). }
\label{fig:sim_vs_crb}\vspace{-0.5cm}
\end{figure}
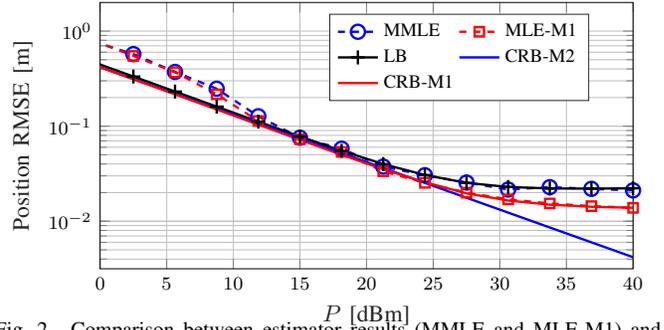

\subsection{Estimation Results vs. Bounds}
\label{sec:estimation_results_vs_bounds}
We first evaluate the position estimation performance of the two estimators (MLE \eqref{eq:MLE_perfect_HWI} and MMLE \eqref{eq:MLE_mismatched}, where observations come from the hardware-impaired model M1). The simulation results are compared with three different lower bounds, namely, LB (the lower bound of using the M2 to process the data from the M1), CRB-M1, and CRB-M2. Note that the average transmission power $P$ is calculated without considering the nonlinearity of the power amplifier (calculated before the PA). Fig.~\ref{fig:sim_vs_crb} shows the positioning errors obtained by the estimators, along with the corresponding theoretical bounds, with respect to $P$. From the figure, we can see that at low transmit powers, the LB and CRBs coincide, implying that the HWIs are not the main source of error. At higher transmit powers (after 20 dBm), LB deviates from the CRBs and the positioning performance is thus more severely affected by HWIs. The MMLE closely follows the LB, indicating the validity of the LB. In terms of CRB-M1, we observe the bound converges to a certain value after \unit[30]{dBm}, due to the PAN, while the CRB-M2 keeps a fixed downward trend.

\begin{figure} 
\begin{minipage}[b]{0.49\linewidth}
\centering
\setlength\fheight{2.3 cm} 
\setlength\fwidth{3 cm}
%
%
\definecolor{mycolor1}{rgb}{0.00000,1.00000,1.00000}%
\definecolor{mycolor2}{rgb}{1.00000,0.00000,1.00000}%
\newcommand\ms{0.8}
\newcommand\lw{0.3}

\pgfplotsset{every tick label/.append style={font=\tiny}}

\begin{tikzpicture}

\begin{axis}[%
width=0.969\fwidth,
height=\fheight,
at={(0\fwidth,0\fheight)},
scale only axis,
xmin=0,
xmax=40,
xlabel style={font=\tiny, yshift=1 ex},
xlabel={$P$ [dBm]},
ymode=log,
ymin=0.01,
ymax=31.6227766016838,
ylabel style={font=\tiny, yshift=-1.5 ex},
ylabel={AEB [$^\circ$]},
axis background/.style={fill=white},
xmajorgrids,
ymajorgrids,
legend style={font=\footnotesize, at={(-0.13, 1.2)}, anchor=south west, legend cell align=left, align=left, draw=white!5!black, legend columns=2}
]

\addplot [color=blue, line width=\lw pt, mark size=\ms pt, mark=o, mark options={solid, blue}]
  table[row sep=crcr]{%
-10	14.0917048781501\\
-6.42857142857144	9.34104514889365\\
-2.85714285714283	6.19201058847074\\
0.714285714285722	4.10466449384721\\
4.28571428571428	2.72110832695702\\
7.85714285714283	1.80411600147892\\
11.4285714285714	1.19645863621449\\
15	0.793945192244335\\
18.5714285714286	0.527555337499391\\
22.1428571428572	0.35159749417092\\
25.7142857142857	0.235852654227369\\
29.2857142857143	0.160340817838088\\
32.8571428571428	0.111802062057083\\
36.4285714285714	0.0813168586098651\\
40	0.0627422637867024\\
};
\addlegendentry{LB (Average)}

\addplot [name path = A, color=mycolor1, line width=\lw pt, forget plot]
  table[row sep=crcr]{%
-10	14.1414165796592\\
-6.42857142857144	9.37392957748823\\
-2.85714285714283	6.21370614557517\\
0.714285714285722	4.11904584580241\\
4.28571428571428	2.73065633045825\\
7.85714285714283	1.8104678852392\\
11.4285714285714	1.20070390411126\\
15	0.801841669319465\\
18.5714285714286	0.542082813139118\\
22.1428571428572	0.374782464965034\\
25.7142857142857	0.270299749428476\\
29.2857142857143	0.208438525204439\\
32.8571428571428	0.174452650855602\\
36.4285714285714	0.157212916642782\\
40	0.149008406760782\\
};

\addplot [name path = B, color=mycolor1, line width=\lw pt, forget plot]
  table[row sep=crcr]{%
-10	14.0222179108657\\
-6.42857142857144	9.29491204519102\\
-2.85714285714283	6.16132132405088\\
0.714285714285722	4.08415707114883\\
4.28571428571428	2.70726657503995\\
7.85714285714283	1.7945667802477\\
11.4285714285714	1.18956524463789\\
15	0.788527784919651\\
18.5714285714286	0.522692107273196\\
22.1428571428572	0.346477753984756\\
25.7142857142857	0.229670831455779\\
29.2857142857143	0.15224350503989\\
32.8571428571428	0.100920005976756\\
36.4285714285714	0.0669003544293331\\
40	0.0443515057639315\\
};

\addplot[fill = mycolor1, area legend] fill between[of=A and B];
\addlegendentry{LB (Multi-Realization)}
    
\addplot [color=mycolor2, dashed, line width=\lw pt, mark size=\ms pt, mark=square, mark options={solid, mycolor2}]
  table[row sep=crcr]{%
-10	13.877072190783\\
-6.42857142857144	9.19869923079369\\
-2.85714285714283	6.09754466758524\\
0.714285714285722	4.04188136174003\\
4.28571428571428	2.67924317622936\\
7.85714285714283	1.77599077135735\\
11.4285714285714	1.1772515641471\\
15	0.780365116552703\\
18.5714285714286	0.517280871547197\\
22.1428571428572	0.342890134877723\\
25.7142857142857	0.227291692122306\\
29.2857142857143	0.150664915822802\\
32.8571428571428	0.0998713003890923\\
36.4285714285714	0.0662017204664891\\
40	0.0438831553774568\\
};
\addlegendentry{CRB-M1 (Average)}

\addplot [name path = C, color=mycolor2, line width=\lw pt, forget plot]
  table[row sep=crcr]{%
-10	13.877072190783\\
-6.42857142857144	9.1986992307937\\
-2.85714285714283	6.09754466758524\\
0.714285714285722	4.04188136174003\\
4.28571428571428	2.67924317622936\\
7.85714285714283	1.77599077135735\\
11.4285714285714	1.1772515641471\\
15	0.780365116552703\\
18.5714285714286	0.517280871547197\\
22.1428571428572	0.342890134877723\\
25.7142857142857	0.227291692122306\\
29.2857142857143	0.150664915822802\\
32.8571428571428	0.0998713003890923\\
36.4285714285714	0.0662017204664892\\
40	0.0438831553774568\\
};
\addplot [name path = D, color=mycolor2, line width=\lw pt, forget plot]
  table[row sep=crcr]{%
-10	13.8770721907829\\
-6.42857142857144	9.19869923079369\\
-2.85714285714283	6.09754466758523\\
0.714285714285722	4.04188136174003\\
4.28571428571428	2.67924317622936\\
7.85714285714283	1.77599077135735\\
11.4285714285714	1.1772515641471\\
15	0.780365116552702\\
18.5714285714286	0.517280871547197\\
22.1428571428572	0.342890134877723\\
25.7142857142857	0.227291692122306\\
29.2857142857143	0.150664915822802\\
32.8571428571428	0.0998713003890922\\
36.4285714285714	0.0662017204664891\\
40	0.0438831553774568\\
};

\addplot[fill = mycolor2, area legend] fill between[of= C and D];
\addlegendentry{CRB-M1 (Multi-Realization)}

\addplot [color=black, line width=\lw pt]
  table[row sep=crcr]{%
-10	13.8770721907829\\
-6.42857142857144	9.19869923079369\\
-2.85714285714283	6.09754466758523\\
0.714285714285722	4.04188136174003\\
4.28571428571428	2.67924317622936\\
7.85714285714283	1.77599077135735\\
11.4285714285714	1.1772515641471\\
15	0.780365116552702\\
18.5714285714286	0.517280871547197\\
22.1428571428572	0.342890134877723\\
25.7142857142857	0.227291692122306\\
29.2857142857143	0.150664915822802\\
32.8571428571428	0.0998713003890922\\
36.4285714285714	0.0662017204664891\\
40	0.0438831553774568\\
};
\addlegendentry{CRB-M2}

\end{axis}

\begin{axis}[%
width=1.25\fwidth,
height=1.25\fheight,
at={(-0.163\fwidth,-0.156\fheight)},
scale only axis,
xmin=0,
xmax=1,
ymin=0,
ymax=1,
axis line style={draw=none},
ticks=none,
axis x line*=bottom,
axis y line*=left
]
\end{axis}
\end{tikzpicture}%
\vspace{-0.5cm}
\end{minipage}
\hfill
\begin{minipage}[b]{0.49\linewidth}
\centering
\setlength\fheight{2.3 cm} 
\setlength\fwidth{3 cm}
%
%
\definecolor{mycolor1}{rgb}{0.00000,1.00000,1.00000}%
\definecolor{mycolor2}{rgb}{1.00000,0.00000,1.00000}%
\newcommand\ms{0.8}
\newcommand\lw{0.3}

\pgfplotsset{every tick label/.append style={font=\tiny}}

\begin{tikzpicture}

\begin{axis}[%
width=0.969\fwidth,
height=\fheight,
at={(0\fwidth,0\fheight)},
scale only axis,
xmin=0,
xmax=40,
xlabel style={font=\tiny, yshift=1 ex},
xlabel={$P$ [dBm]},
ymode=log,
ymin=0.000316227766016838,
ymax=1,
yminorticks=true,
ylabel style={font=\tiny, yshift=-1.5 ex},
ylabel={DEB [m]},
axis background/.style={fill=white},
xmajorgrids,
ymajorgrids,
legend style={font=\tiny, at={(0.03,1.23)}, anchor=south west, legend cell align=left, align=left, draw=white!5!black}
]

\addplot [color=blue, line width=\lw pt, mark size=\ms pt, mark=o, mark options={solid, blue}]
  table[row sep=crcr]{%
-10	0.234078451053771\\
-6.42857142857144	0.15516444078073\\
-2.85714285714283	0.102855084516361\\
0.714285714285722	0.0681813663580031\\
4.28571428571428	0.0451980948644631\\
7.85714285714283	0.0299645340551678\\
11.4285714285714	0.0198687196192976\\
15	0.0131796047307712\\
18.5714285714286	0.00875023933431008\\
22.1428571428572	0.0058210676932505\\
25.7142857142857	0.00388955174647267\\
29.2857142857143	0.00262359913341078\\
32.8571428571428	0.0018036684222522\\
36.4285714285714	0.00128354470360521\\
40	0.000963855445942191\\
};

\addplot [name path = A, color=mycolor1, line width=\lw pt, forget plot]
  table[row sep=crcr]{%
-10	0.234939363720538\\
-6.42857142857144	0.155734396656378\\
-2.85714285714283	0.103231808836778\\
0.714285714285722	0.0684294527502525\\
4.28571428571428	0.0453600807567971\\
7.85714285714283	0.0300681946913009\\
11.4285714285714	0.0199379454535355\\
15	0.013252453142209\\
18.5714285714286	0.00885532365470277\\
22.1428571428572	0.00597516831988176\\
25.7142857142857	0.00411511574914339\\
29.2857142857143	0.00294743482408716\\
32.8571428571428	0.00225027775485975\\
36.4285714285714	0.0018632072063893\\
40	0.00166491116651767\\
};

\addplot [name path = B, color=mycolor1, line width=\lw pt, forget plot]
  table[row sep=crcr]{%
-10	0.233152262970797\\
-6.42857142857144	0.154549783572399\\
-2.85714285714283	0.102446568445211\\
0.714285714285722	0.0679089469288588\\
4.28571428571428	0.0450150630439467\\
7.85714285714283	0.0298395079961053\\
11.4285714285714	0.0197802655576238\\
15	0.0131125676855073\\
18.5714285714286	0.00869316398057939\\
22.1428571428572	0.00576429864901201\\
25.7142857142857	0.00382378439175298\\
29.2857142857143	0.00253889506007138\\
32.8571428571428	0.00168426193018226\\
36.4285714285714	0.00111703892421825\\
40	0.000741343783430686\\
};

\addplot[fill = mycolor1, area legend] fill between[of=A and B];

\addplot [color=mycolor2, dashed, line width=\lw pt, mark size=\ms pt, mark=square, mark options={solid, mycolor2}]
  table[row sep=crcr]{%
-10	0.230510309256065\\
-6.42857142857144	0.152798441579927\\
-2.85714285714283	0.101285551282301\\
0.714285714285722	0.0671391854032325\\
4.28571428571428	0.044504573056486\\
7.85714285714283	0.0295007604135266\\
11.4285714285714	0.0195551783829427\\
15	0.0129625472777093\\
18.5714285714286	0.00859248781250763\\
22.1428571428572	0.00569570511307243\\
25.7142857142857	0.00377551384918543\\
29.2857142857143	0.00250267605896151\\
32.8571428571428	0.00165894967050655\\
36.4285714285714	0.00109966849261977\\
40	0.000728937601398958\\
};

\addplot [name path = C, color=mycolor2, line width=\lw pt, forget plot]
  table[row sep=crcr]{%
-10	0.230510309256065\\
-6.42857142857144	0.152798441579927\\
-2.85714285714283	0.101285551282301\\
0.714285714285722	0.0671391854032326\\
4.28571428571428	0.044504573056486\\
7.85714285714283	0.0295007604135266\\
11.4285714285714	0.0195551783829427\\
15	0.0129625472777093\\
18.5714285714286	0.00859248781250764\\
22.1428571428572	0.00569570511307244\\
25.7142857142857	0.00377551384918544\\
29.2857142857143	0.00250267605896151\\
32.8571428571428	0.00165894967050655\\
36.4285714285714	0.00109966849261977\\
40	0.000728937601398959\\
};
\addplot [name path = D, color=mycolor2, line width=\lw pt, forget plot]
  table[row sep=crcr]{%
-10	0.230510309256065\\
-6.42857142857144	0.152798441579927\\
-2.85714285714283	0.101285551282301\\
0.714285714285722	0.0671391854032325\\
4.28571428571428	0.044504573056486\\
7.85714285714283	0.0295007604135265\\
11.4285714285714	0.0195551783829427\\
15	0.0129625472777093\\
18.5714285714286	0.00859248781250762\\
22.1428571428572	0.00569570511307243\\
25.7142857142857	0.00377551384918543\\
29.2857142857143	0.0025026760589615\\
32.8571428571428	0.00165894967050655\\
36.4285714285714	0.00109966849261977\\
40	0.000728937601398957\\
};

\addplot[fill = mycolor2, area legend] fill between[of= C and D];

\addplot [color=black, line width=\lw pt]
  table[row sep=crcr]{%
-10	0.230510309256065\\
-6.42857142857144	0.152798441579927\\
-2.85714285714283	0.101285551282301\\
0.714285714285722	0.0671391854032326\\
4.28571428571428	0.044504573056486\\
7.85714285714283	0.0295007604135266\\
11.4285714285714	0.0195551783829427\\
15	0.0129625472777093\\
18.5714285714286	0.00859248781250764\\
22.1428571428572	0.00569570511307243\\
25.7142857142857	0.00377551384918544\\
29.2857142857143	0.00250267605896151\\
32.8571428571428	0.00165894967050655\\
36.4285714285714	0.00109966849261977\\
40	0.000728937601398958\\
};

\end{axis}

\begin{axis}[%
width=1.25\fwidth,
height=1.25\fheight,
at={(-0.163\fwidth,-0.156\fheight)},
scale only axis,
xmin=0,
xmax=1,
ymin=0,
ymax=1,
axis line style={draw=none},
ticks=none,
axis x line*=bottom,
axis y line*=left
]
\end{axis}
\end{tikzpicture}%
\vspace{-0.5cm}
\end{minipage}
\centerline{(a) PN ($\sigma_\text{PN}=10^\circ$)} 
\medskip
\hfill
\begin{minipage}[b]{0.49\linewidth}
\centering
\setlength\fheight{2.3 cm} 
\setlength\fwidth{3 cm}
%
%
\definecolor{mycolor1}{rgb}{0.00000,1.00000,1.00000}%
\definecolor{mycolor2}{rgb}{1.00000,0.00000,1.00000}%
\newcommand\ms{0.8}
\newcommand\lw{0.3}

\pgfplotsset{every tick label/.append style={font=\tiny}}

\begin{tikzpicture}

\begin{axis}[%
width=0.969\fwidth,
height=\fheight,
at={(0\fwidth,0\fheight)},
scale only axis,
xmin=0,
xmax=40,
xlabel style={font=\tiny, yshift=1 ex},
xlabel={$P$ [dBm]},
ymode=log,
ymin=0.01,
ymax=31.6227766016838,
ylabel style={font=\tiny, yshift=-1.5 ex},
ylabel={AEB [$^\circ$]},
axis background/.style={fill=white},
xmajorgrids,
ymajorgrids,
legend style={font=\footnotesize, at={(-0.13, 1.2)}, anchor=south west, legend cell align=left, align=left, draw=white!5!black, legend columns=2}
]

\addplot [color=blue, line width=\lw pt, mark size=\ms pt, mark=o, mark options={solid, blue}]
table[row sep=crcr]{%
-10	14.1981623682458\\
-6.42857142857144	9.41229765424204\\
-2.85714285714283	6.24027439072502\\
0.714285714285722	4.13821240585183\\
4.28571428571428	2.74568387286313\\
7.85714285714283	1.8239041240582\\
11.4285714285714	1.21475930365763\\
15	0.813615497929585\\
18.5714285714286	0.551209268494806\\
22.1428571428572	0.381536084469337\\
25.7142857142857	0.273792316976195\\
29.2857142857143	0.20709465236627\\
32.8571428571428	0.167087607894878\\
36.4285714285714	0.143888401315548\\
40	0.130865675009575\\
};

\addplot [name path = A, color=mycolor1, line width=\lw pt, forget plot]
table[row sep=crcr]{%
-10	18.1911965663179\\
-6.42857142857144	12.0593979753218\\
-2.85714285714283	7.99531591712596\\
0.714285714285722	5.30211822542655\\
4.28571428571428	3.51802477885039\\
7.85714285714283	2.33712732636737\\
11.4285714285714	1.55692840082262\\
15	1.04358857346149\\
18.5714285714286	0.760866929513919\\
22.1428571428572	0.648002814173523\\
25.7142857142857	0.591640730067139\\
29.2857142857143	0.565100493055121\\
32.8571428571428	0.553036161381348\\
36.4285714285714	0.54765109134925\\
40	0.54526808944786\\
};

\addplot [name path = B, color=mycolor1, line width=\lw pt, forget plot]
table[row sep=crcr]{%
-10	13.6685432132596\\
-6.42857142857144	9.06120549456168\\
-2.85714285714283	6.00751120888251\\
0.714285714285722	3.98387044758936\\
4.28571428571428	2.64330646854456\\
7.85714285714283	1.75596061596407\\
11.4285714285714	1.16568897378517\\
15	0.774724304372427\\
18.5714285714286	0.51475376727444\\
22.1428571428572	0.342006980384085\\
25.7142857142857	0.22694107587982\\
29.2857142857143	0.150553763000742\\
32.8571428571428	0.0999219250230977\\
36.4285714285714	0.0662481486886784\\
40	0.0439333416424818\\
};

\addplot[fill = mycolor1, area legend] fill between[of=A and B];
    
\addplot [color=mycolor2, dashed, line width=\lw pt, mark size=\ms pt, mark=square, mark options={solid, mycolor2}]
table[row sep=crcr]{%
-10	13.8770721907829\\
-6.42857142857144	9.19869923079369\\
-2.85714285714283	6.09754466758524\\
0.714285714285722	4.04188136174003\\
4.28571428571428	2.67924317622936\\
7.85714285714283	1.77599077135735\\
11.4285714285714	1.1772515641471\\
15	0.780365116552703\\
18.5714285714286	0.517280871547197\\
22.1428571428572	0.342890134877723\\
25.7142857142857	0.227291692122306\\
29.2857142857143	0.150664915822802\\
32.8571428571428	0.0998713003890923\\
36.4285714285714	0.0662017204664891\\
40	0.0438831553774568\\
};

\addplot [name path = C, color=mycolor2, line width=\lw pt, forget plot]
table[row sep=crcr]{%
-10	13.877072190783\\
-6.42857142857144	9.1986992307937\\
-2.85714285714283	6.09754466758524\\
0.714285714285722	4.04188136174003\\
4.28571428571428	2.67924317622936\\
7.85714285714283	1.77599077135735\\
11.4285714285714	1.1772515641471\\
15	0.780365116552703\\
18.5714285714286	0.517280871547197\\
22.1428571428572	0.342890134877723\\
25.7142857142857	0.227291692122306\\
29.2857142857143	0.150664915822802\\
32.8571428571428	0.0998713003890923\\
36.4285714285714	0.0662017204664892\\
40	0.0438831553774568\\
};
\addplot [name path = D, color=mycolor2, line width=\lw pt, forget plot]
table[row sep=crcr]{%
-10	13.8770721907829\\
-6.42857142857144	9.19869923079369\\
-2.85714285714283	6.09754466758523\\
0.714285714285722	4.04188136174003\\
4.28571428571428	2.67924317622936\\
7.85714285714283	1.77599077135735\\
11.4285714285714	1.1772515641471\\
15	0.780365116552702\\
18.5714285714286	0.517280871547197\\
22.1428571428572	0.342890134877723\\
25.7142857142857	0.227291692122306\\
29.2857142857143	0.150664915822802\\
32.8571428571428	0.0998713003890922\\
36.4285714285714	0.0662017204664891\\
40	0.0438831553774568\\
};

\addplot[fill = mycolor2, area legend] fill between[of= C and D];

\addplot [color=black, line width=\lw pt]
table[row sep=crcr]{%
-10	13.8770721907829\\
-6.42857142857144	9.19869923079369\\
-2.85714285714283	6.09754466758523\\
0.714285714285722	4.04188136174003\\
4.28571428571428	2.67924317622936\\
7.85714285714283	1.77599077135735\\
11.4285714285714	1.1772515641471\\
15	0.780365116552702\\
18.5714285714286	0.517280871547197\\
22.1428571428572	0.342890134877723\\
25.7142857142857	0.227291692122306\\
29.2857142857143	0.150664915822802\\
32.8571428571428	0.0998713003890922\\
36.4285714285714	0.0662017204664891\\
40	0.0438831553774568\\
};

\end{axis}

\begin{axis}[%
width=1.25\fwidth,
height=1.25\fheight,
at={(-0.163\fwidth,-0.156\fheight)},
scale only axis,
xmin=0,
xmax=1,
ymin=0,
ymax=1,
axis line style={draw=none},
ticks=none,
axis x line*=bottom,
axis y line*=left
]
\end{axis}
\end{tikzpicture}%
\vspace{-0.5cm}
\end{minipage}
\hfill
\begin{minipage}[b]{0.49\linewidth}
  \centering
\setlength\fheight{2.3 cm} 
\setlength\fwidth{3 cm}
%
%
\definecolor{mycolor1}{rgb}{0.00000,1.00000,1.00000}%
\definecolor{mycolor2}{rgb}{1.00000,0.00000,1.00000}%
\newcommand\ms{0.8}
\newcommand\lw{0.3}

\pgfplotsset{every tick label/.append style={font=\tiny}}

\begin{tikzpicture}

\begin{axis}[%
width=0.969\fwidth,
height=\fheight,
at={(0\fwidth,0\fheight)},
scale only axis,
xmin=0,
xmax=40,
xlabel style={font=\tiny, yshift=1 ex},
xlabel={$P$ [dBm]},
ymode=log,
ymin=0.000316227766016838,
ymax=1,
yminorticks=true,
ylabel style={font=\tiny, yshift=-1.5 ex},
ylabel={DEB [m]},
axis background/.style={fill=white},
xmajorgrids,
ymajorgrids,
legend style={font=\tiny, at={(0.03,1.23)}, anchor=south west, legend cell align=left, align=left, draw=white!5!black}
]

\addplot [color=blue, line width=\lw pt, mark size=\ms pt, mark=o, mark options={solid, blue}]
table[row sep=crcr]{%
-10	0.235994524388121\\
-6.42857142857144	0.156433926228873\\
-2.85714285714283	0.10369564937083\\
0.714285714285722	0.0687371348154387\\
4.28571428571428	0.0455643598227866\\
7.85714285714283	0.0302040966130988\\
11.4285714285714	0.020022658604088\\
15	0.013274325618863\\
18.5714285714286	0.00880202025999387\\
22.1428571428572	0.00583891200451859\\
25.7142857142857	0.00387692816819852\\
29.2857142857143	0.00257961474699167\\
32.8571428571428	0.00172437197542381\\
36.4285714285714	0.00116407860415396\\
40	0.000801400788995889\\
};

\addplot [name path = A, color=mycolor1, line width=\lw pt, forget plot]
table[row sep=crcr]{%
-10	0.281169779772016\\
-6.42857142857144	0.186380420175209\\
-2.85714285714283	0.123548038398001\\
0.714285714285722	0.0818993298151719\\
4.28571428571428	0.0542931639795266\\
7.85714285714283	0.0359961587450223\\
11.4285714285714	0.0238710885745095\\
15	0.0158389687354437\\
18.5714285714286	0.0105225771525685\\
22.1428571428572	0.00701026955274211\\
25.7142857142857	0.00469951897752329\\
29.2857142857143	0.00319313279762415\\
32.8571428571428	0.0022297842517273\\
36.4285714285714	0.00163598826735061\\
40	0.00129144507084528\\
};

\addplot [name path = B, color=mycolor1, line width=\lw pt, forget plot]
table[row sep=crcr]{%
-10	0.230511023303905\\
-6.42857142857144	0.1527989149768\\
-2.85714285714283	0.10128586519732\\
0.714285714285722	0.067139393660588\\
4.28571428571428	0.0445047113641979\\
7.85714285714283	0.0295008524859768\\
11.4285714285714	0.0195552400069679\\
15	0.012962589019428\\
18.5714285714286	0.00859251682899755\\
22.1428571428572	0.0056957263795301\\
25.7142857142857	0.0037755310119777\\
29.2857142857143	0.00250269206083216\\
32.8571428571428	0.00165896725513163\\
36.4285714285714	0.00109969067500458\\
40	0.000728968184708993\\
};

\addplot[fill = mycolor1, area legend] fill between[of=A and B];

\addplot [color=mycolor2, dashed, line width=\lw pt, mark size=\ms pt, mark=square, mark options={solid, mycolor2}]
  table[row sep=crcr]{%
-10	0.230510309256065\\
-6.42857142857144	0.152798441579927\\
-2.85714285714283	0.101285551282301\\
0.714285714285722	0.0671391854032325\\
4.28571428571428	0.044504573056486\\
7.85714285714283	0.0295007604135266\\
11.4285714285714	0.0195551783829427\\
15	0.0129625472777093\\
18.5714285714286	0.00859248781250763\\
22.1428571428572	0.00569570511307243\\
25.7142857142857	0.00377551384918543\\
29.2857142857143	0.00250267605896151\\
32.8571428571428	0.00165894967050655\\
36.4285714285714	0.00109966849261977\\
40	0.000728937601398958\\
};

\addplot [name path = C, color=mycolor2, line width=\lw pt, forget plot]
table[row sep=crcr]{%
-10	0.230510309256065\\
-6.42857142857144	0.152798441579927\\
-2.85714285714283	0.101285551282301\\
0.714285714285722	0.0671391854032325\\
4.28571428571428	0.044504573056486\\
7.85714285714283	0.0295007604135266\\
11.4285714285714	0.0195551783829427\\
15	0.0129625472777093\\
18.5714285714286	0.00859248781250763\\
22.1428571428572	0.00569570511307243\\
25.7142857142857	0.00377551384918543\\
29.2857142857143	0.0025026760589615\\
32.8571428571428	0.00165894967050655\\
36.4285714285714	0.00109966849261977\\
40	0.000728937601398958\\
};
\addplot [name path = D, color=mycolor2, line width=\lw pt, forget plot]
table[row sep=crcr]{%
-10	0.230510309256065\\
-6.42857142857144	0.152798441579927\\
-2.85714285714283	0.101285551282301\\
0.714285714285722	0.0671391854032326\\
4.28571428571428	0.044504573056486\\
7.85714285714283	0.0295007604135266\\
11.4285714285714	0.0195551783829427\\
15	0.0129625472777093\\
18.5714285714286	0.00859248781250764\\
22.1428571428572	0.00569570511307244\\
25.7142857142857	0.00377551384918544\\
29.2857142857143	0.00250267605896151\\
32.8571428571428	0.00165894967050655\\
36.4285714285714	0.00109966849261977\\
40	0.000728937601398958\\
};

\addplot[fill = mycolor2, area legend] fill between[of= C and D];

\addplot [color=black, line width=\lw pt]
table[row sep=crcr]{%
-10	0.230510309256065\\
-6.42857142857144	0.152798441579927\\
-2.85714285714283	0.101285551282301\\
0.714285714285722	0.0671391854032325\\
4.28571428571428	0.044504573056486\\
7.85714285714283	0.0295007604135265\\
11.4285714285714	0.0195551783829427\\
15	0.0129625472777092\\
18.5714285714286	0.00859248781250763\\
22.1428571428572	0.00569570511307242\\
25.7142857142857	0.00377551384918543\\
29.2857142857143	0.0025026760589615\\
32.8571428571428	0.00165894967050655\\
36.4285714285714	0.00109966849261976\\
40	0.000728937601398957\\
};

\end{axis}

\begin{axis}[%
width=1.25\fwidth,
height=1.25\fheight,
at={(-0.163\fwidth,-0.156\fheight)},
scale only axis,
xmin=0,
xmax=1,
ymin=0,
ymax=1,
axis line style={draw=none},
ticks=none,
axis x line*=bottom,
axis y line*=left
]
\end{axis}
\end{tikzpicture}%
\vspace{-0.5cm}
\end{minipage}
\centerline{(b) CFO ($\sigma_\text{CFO}=0.01$)}\medskip
\hfill
\begin{minipage}[b]{0.49\linewidth}
\centering
\setlength\fheight{2.3 cm} 
\setlength\fwidth{3 cm}
%
%
\definecolor{mycolor1}{rgb}{0.00000,1.00000,1.00000}%
\definecolor{mycolor2}{rgb}{1.00000,0.00000,1.00000}%
\newcommand\ms{0.8}
\newcommand\lw{0.3}

\pgfplotsset{every tick label/.append style={font=\tiny}}

\begin{tikzpicture}

\begin{axis}[%
width=0.969\fwidth,
height=\fheight,
at={(0\fwidth,0\fheight)},
scale only axis,
xmin=0,
xmax=40,
xlabel style={font=\tiny, yshift=1 ex},
xlabel={$P$ [dBm]},
ymode=log,
ymin=0.01,
ymax=31.6227766016838,
ylabel style={font=\tiny, yshift=-1.5 ex},
ylabel={AEB [$^\circ$]},
axis background/.style={fill=white},
xmajorgrids,
ymajorgrids,
legend style={font=\footnotesize, at={(-0.13, 1.2)}, anchor=south west, legend cell align=left, align=left, draw=white!5!black, legend columns=2}
]

\addplot [color=blue, line width=\lw pt, mark size=\ms pt, mark=o, mark options={solid, blue}]
table[row sep=crcr]{%
-10	14.0053766158233\\
-6.42857142857144	9.286621631847\\
-2.85714285714283	6.16015427933356\\
0.714285714285722	4.08989277252169\\
4.28571428571428	2.72081941849018\\
7.85714285714283	1.8180337162897\\
11.4285714285714	1.226261034327\\
15	0.842726256521988\\
18.5714285714286	0.59873805794047\\
22.1428571428572	0.447413253350284\\
25.7142857142857	0.356252717447625\\
29.2857142857143	0.302956066800333\\
32.8571428571428	0.27266665164696\\
36.4285714285714	0.255863443531488\\
40	0.246717974669255\\
};

\addplot [name path = A, color=mycolor1, line width=\lw pt, forget plot]
table[row sep=crcr]{%
-10	15.0362159553695\\
-6.42857142857144	9.96716190372489\\
-2.85714285714283	6.6070876437571\\
0.714285714285722	4.37987139339054\\
4.28571428571428	2.91633152156752\\
7.85714285714283	1.98223678343279\\
11.4285714285714	1.39763369486739\\
15	1.05792225107999\\
18.5714285714286	0.867561555872722\\
22.1428571428572	0.769160595971279\\
25.7142857142857	0.72169308969264\\
29.2857142857143	0.699818588883716\\
32.8571428571428	0.689987730467661\\
36.4285714285714	0.685623505364108\\
40	0.683697067922503\\
};

\addplot [name path = B, color=mycolor1, line width=\lw pt, forget plot]
table[row sep=crcr]{%
-10	13.0867694394009\\
-6.42857142857144	8.68528266220157\\
-2.85714285714283	5.76956117648747\\
0.714285714285722	3.82687232562684\\
4.28571428571428	2.54033920698946\\
7.85714285714283	1.68936255279058\\
11.4285714285714	1.1280026501491\\
15	0.749732317541804\\
18.5714285714286	0.498641119523878\\
22.1428571428572	0.333033791336795\\
25.7142857142857	0.223283328404022\\
29.2857142857143	0.148274184612924\\
32.8571428571428	0.0986871075205124\\
36.4285714285714	0.0660170424858278\\
40	0.0442055331271263\\
};

\addplot[fill = mycolor1, area legend] fill between[of=A and B];
    
\addplot [color=mycolor2, dashed, line width=\lw pt, mark size=\ms pt, mark=square, mark options={solid, mycolor2}]
table[row sep=crcr]{%
-10	13.9414603163868\\
-6.42857142857144	9.24138020797116\\
-2.85714285714283	6.12583662042184\\
0.714285714285722	4.06063525746226\\
4.28571428571428	2.69167457701641\\
7.85714285714283	1.78423117792765\\
11.4285714285714	1.18271388505583\\
15	0.783985926940564\\
18.5714285714286	0.519680999273836\\
22.1428571428572	0.344481108302759\\
25.7142857142857	0.228346301179598\\
29.2857142857143	0.151363984862058\\
32.8571428571428	0.100334692504177\\
36.4285714285714	0.0665088893443322\\
40	0.0440867685086341\\
};

\addplot [name path = C, color=mycolor2, line width=\lw pt, forget plot]
table[row sep=crcr]{%
-10	14.6556599663493\\
-6.42857142857144	9.71480195575931\\
-2.85714285714283	6.43965384406592\\
0.714285714285722	4.26865537972277\\
4.28571428571428	2.82956494123158\\
7.85714285714283	1.87563460725351\\
11.4285714285714	1.24330250515324\\
15	0.824148324701602\\
18.5714285714286	0.546303460576349\\
22.1428571428572	0.362128347643915\\
25.7142857142857	0.240044132301421\\
29.2857142857143	0.159118129876432\\
32.8571428571428	0.105474685061582\\
36.4285714285714	0.0699160378360352\\
40	0.0463452661066105\\
};
\addplot [name path = D, color=mycolor2, line width=\lw pt, forget plot]
table[row sep=crcr]{%
-10	13.0065752091302\\
-6.42857142857144	8.62167262132949\\
-2.85714285714283	5.7150508565239\\
0.714285714285722	3.7883375682639\\
4.28571428571428	2.51117652168169\\
7.85714285714283	1.66458437491758\\
11.4285714285714	1.1034035709143\\
15	0.731413473929021\\
18.5714285714286	0.484832280723577\\
22.1428571428572	0.321380927218789\\
25.7142857142857	0.213033876840591\\
29.2857142857143	0.141213833298937\\
32.8571428571428	0.0936064583282298\\
36.4285714285714	0.062048942628771\\
40	0.0411304022191323\\
};

\addplot[fill = mycolor2, area legend] fill between[of= C and D];

\addplot [color=black, line width=\lw pt]
table[row sep=crcr]{%
-10	13.8770721907829\\
-6.42857142857144	9.19869923079369\\
-2.85714285714283	6.09754466758523\\
0.714285714285722	4.04188136174003\\
4.28571428571428	2.67924317622936\\
7.85714285714283	1.77599077135735\\
11.4285714285714	1.1772515641471\\
15	0.780365116552702\\
18.5714285714286	0.517280871547197\\
22.1428571428572	0.342890134877723\\
25.7142857142857	0.227291692122306\\
29.2857142857143	0.150664915822802\\
32.8571428571428	0.0998713003890922\\
36.4285714285714	0.0662017204664891\\
40	0.0438831553774568\\
};

\end{axis}

\begin{axis}[%
width=1.25\fwidth,
height=1.25\fheight,
at={(-0.163\fwidth,-0.156\fheight)},
scale only axis,
xmin=0,
xmax=1,
ymin=0,
ymax=1,
axis line style={draw=none},
ticks=none,
axis x line*=bottom,
axis y line*=left
]
\end{axis}
\end{tikzpicture}%
\vspace{-0.5cm}
\end{minipage}
\hfill
\begin{minipage}[b]{0.49\linewidth}
\centering
\setlength\fheight{2.3 cm} 
\setlength\fwidth{3 cm}
%
%
\definecolor{mycolor1}{rgb}{0.00000,1.00000,1.00000}%
\definecolor{mycolor2}{rgb}{1.00000,0.00000,1.00000}%
\newcommand\ms{0.8}
\newcommand\lw{0.3}

\pgfplotsset{every tick label/.append style={font=\tiny}}

\begin{tikzpicture}

\begin{axis}[%
width=0.969\fwidth,
height=\fheight,
at={(0\fwidth,0\fheight)},
scale only axis,
xmin=0,
xmax=40,
xlabel style={font=\tiny, yshift=1 ex},
xlabel={$P$ [dBm]},
ymode=log,
ymin=0.000316227766016838,
ymax=1,
yminorticks=true,
ylabel style={font=\tiny, yshift=-1.5 ex},
ylabel={DEB [m]},
axis background/.style={fill=white},
xmajorgrids,
ymajorgrids,
legend style={font=\tiny, at={(0.03,1.23)}, anchor=south west, legend cell align=left, align=left, draw=white!5!black}
]

\addplot [color=blue, line width=\lw pt, mark size=\ms pt, mark=o, mark options={solid, blue}]
table[row sep=crcr]{%
-10	0.232248951969598\\
-6.42857142857144	0.153950955973247\\
-2.85714285714283	0.102049548654413\\
0.714285714285722	0.0676456615260814\\
4.28571428571428	0.0448403688442279\\
7.85714285714283	0.0297234517528169\\
11.4285714285714	0.0197029481478315\\
15	0.0130607321851974\\
18.5714285714286	0.00865792261617167\\
22.1428571428572	0.00573960888185723\\
25.7142857142857	0.00380541283568942\\
29.2857142857143	0.00252369159634289\\
32.8571428571428	0.00167466794352929\\
36.4285714285714	0.00111272998791422\\
40	0.000741421452578347\\
};

\addplot [name path = A, color=mycolor1, line width=\lw pt, forget plot]
table[row sep=crcr]{%
-10	0.250350424034827\\
-6.42857142857144	0.165949869947171\\
-2.85714285714283	0.110003250623246\\
0.714285714285722	0.0729179013806075\\
4.28571428571428	0.0483351302336984\\
7.85714285714283	0.032039950049669\\
11.4285714285714	0.0212383725350765\\
15	0.0140783481398764\\
18.5714285714286	0.00933221174333006\\
22.1428571428572	0.00618618593350958\\
25.7142857142857	0.00411328519774402\\
29.2857142857143	0.00275585860943466\\
32.8571428571428	0.00187020346754608\\
36.4285714285714	0.00130284659762731\\
40	0.000952041977969193\\
};

\addplot [name path = B, color=mycolor1, line width=\lw pt, forget plot]
  table[row sep=crcr]{%
-10	0.216214240405521\\
-6.42857142857144	0.143322007781945\\
-2.85714285714283	0.0950039134865909\\
0.714285714285722	0.0629752875560212\\
4.28571428571428	0.0417444689657919\\
7.85714285714283	0.0276711998034304\\
11.4285714285714	0.0183424629162504\\
15	0.0121587435174489\\
18.5714285714286	0.00805977474124297\\
22.1428571428572	0.00534274319833642\\
25.7142857142857	0.00354178386785784\\
29.2857142857143	0.00234810238940998\\
32.8571428571428	0.00155702881737126\\
36.4285714285714	0.0010329245371035\\
40	0.000685924965120125\\
};

\addplot[fill = mycolor1, area legend] fill between[of=A and B];

\addplot [color=mycolor2, dashed, line width=\lw pt, mark size=\ms pt, mark=square, mark options={solid, mycolor2}]
table[row sep=crcr]{%
-10	0.231735507989138\\
-6.42857142857144	0.153610589451506\\
-2.85714285714283	0.101823899998722\\
0.714285714285722	0.0674960407871027\\
4.28571428571428	0.0447411218976235\\
7.85714285714283	0.029657561618644\\
11.4285714285714	0.0196591172473567\\
15	0.0130314452656271\\
18.5714285714286	0.00863815824354305\\
22.1428571428572	0.00572597868613307\\
25.7142857142857	0.00379558130213212\\
29.2857142857143	0.00251597817784126\\
32.8571428571428	0.00166776725025427\\
36.4285714285714	0.0011055134044951\\
40	0.000732812019981822\\
};

\addplot [name path = C, color=mycolor2, line width=\lw pt, forget plot]
table[row sep=crcr]{%
-10	0.24871063370664\\
-6.42857142857144	0.164862896403106\\
-2.85714285714283	0.109282720265513\\
0.714285714285722	0.0724402713357003\\
4.28571428571428	0.0480185055646523\\
7.85714285714283	0.0318300419662592\\
11.4285714285714	0.021099189982281\\
15	0.0139860267347523\\
18.5714285714286	0.0092709219638042\\
22.1428571428572	0.00614541897345149\\
25.7142857142857	0.00407361581800662\\
29.2857142857143	0.00270027900528867\\
32.8571428571428	0.00178993479801706\\
36.4285714285714	0.00118649464550788\\
40	0.000786492080816826\\
};
\addplot [name path = D, color=mycolor2, line width=\lw pt, forget plot]
 table[row sep=crcr]{%
-10	0.216984945135719\\
-6.42857142857144	0.143832879188988\\
-2.85714285714283	0.0953425461054641\\
0.714285714285722	0.0631997436825874\\
4.28571428571428	0.0418932340775389\\
7.85714285714283	0.0277697813188918\\
11.4285714285714	0.0184077637231766\\
15	0.0122019601593975\\
18.5714285714286	0.00808831718890788\\
22.1428571428572	0.00536150537239686\\
25.7142857142857	0.00355398276141566\\
29.2857142857143	0.00235582967676727\\
32.8571428571428	0.00156160956271118\\
36.4285714285714	0.0010351446245882\\
40	0.000686166644595544\\
};

\addplot[fill = mycolor2, area legend] fill between[of= C and D];

\addplot [color=black, line width=\lw pt]
table[row sep=crcr]{%
-10	0.230510309256065\\
-6.42857142857144	0.152798441579927\\
-2.85714285714283	0.101285551282301\\
0.714285714285722	0.0671391854032326\\
4.28571428571428	0.044504573056486\\
7.85714285714283	0.0295007604135266\\
11.4285714285714	0.0195551783829427\\
15	0.0129625472777093\\
18.5714285714286	0.00859248781250764\\
22.1428571428572	0.00569570511307243\\
25.7142857142857	0.00377551384918544\\
29.2857142857143	0.00250267605896151\\
32.8571428571428	0.00165894967050655\\
36.4285714285714	0.00109966849261977\\
40	0.000728937601398958\\
};

\end{axis}

\begin{axis}[%
width=1.25\fwidth,
height=1.25\fheight,
at={(-0.163\fwidth,-0.156\fheight)},
scale only axis,
xmin=0,
xmax=1,
ymin=0,
ymax=1,
axis line style={draw=none},
ticks=none,
axis x line*=bottom,
axis y line*=left
]
\end{axis}
\end{tikzpicture}%
\vspace{-0.5cm}
\end{minipage}
\centerline{(c) MC ($\sigma_\text{MC}=0.02$)} 
\medskip
\hfill
\begin{minipage}[b]{0.49\linewidth}
\centering
\setlength\fheight{2.3 cm} 
\setlength\fwidth{3 cm}
%
%
\definecolor{mycolor1}{rgb}{0.00000,1.00000,1.00000}%
\definecolor{mycolor2}{rgb}{1.00000,0.00000,1.00000}%
\newcommand\ms{0.8}
\newcommand\lw{0.3}

\pgfplotsset{every tick label/.append style={font=\tiny}}

\begin{tikzpicture}

\begin{axis}[%
width=0.969\fwidth,
height=\fheight,
at={(0\fwidth,0\fheight)},
scale only axis,
xmin=0,
xmax=40,
xlabel style={font=\tiny, yshift=1 ex},
xlabel={$P$ [dBm]},
ymode=log,
ymin=0.01,
ymax=31.6227766016838,
ylabel style={font=\tiny, yshift=-1.5 ex},
ylabel={AEB [$^\circ$]},
axis background/.style={fill=white},
xmajorgrids,
ymajorgrids,
legend style={font=\footnotesize, at={(-0.13, 1.2)}, anchor=south west, legend cell align=left, align=left, draw=white!5!black, legend columns=2}
]

\addplot [color=blue, line width=\lw pt, mark size=\ms pt, mark=o, mark options={solid, blue}]
table[row sep=crcr]{%
-10	13.8237863084718\\
-6.42857142857144	9.15096923499022\\
-2.85714285714283	6.05373982181038\\
0.714285714285722	4.00103988359725\\
4.28571428571428	2.64090251709951\\
7.85714285714283	1.7400981931945\\
11.4285714285714	1.14413478244431\\
15	0.750750769255665\\
18.5714285714286	0.492359283759677\\
22.1428571428572	0.324461220257669\\
25.7142857142857	0.21818275089981\\
29.2857142857143	0.156393716476203\\
32.8571428571428	0.141228615252915\\
36.4285714285714	0.141228615252915\\
40	0.141228615252914\\
};

\addplot [name path = A, color=mycolor1, line width=\lw pt, forget plot]
  table[row sep=crcr]{%
-10	19.3263348299784\\
-6.42857142857144	12.7935061717402\\
-2.85714285714283	8.46342674569416\\
0.714285714285722	5.59365102534245\\
4.28571428571428	3.69211192649289\\
7.85714285714283	2.43274306823649\\
11.4285714285714	1.5995568365081\\
15	1.04958659058603\\
18.5714285714286	0.68834255407698\\
22.1428571428572	0.45361278322135\\
25.7142857142857	0.305030243084077\\
29.2857142857143	0.21864612650092\\
32.8571428571428	0.197444567287567\\
36.4285714285714	0.197444567287567\\
40	0.197444567287567\\
};

\addplot [name path = B, color=mycolor1, line width=\lw pt, forget plot]
  table[row sep=crcr]{%
-10	8.61582276571573\\
-6.42857142857144	5.70343951388156\\
-2.85714285714283	3.77305813404458\\
0.714285714285722	2.49369092854884\\
4.28571428571428	1.64597075802001\\
7.85714285714283	1.08453482229525\\
11.4285714285714	0.71309424825163\\
15	0.467913451842602\\
18.5714285714286	0.306868193074453\\
22.1428571428572	0.202223928069167\\
25.7142857142857	0.135984734597425\\
29.2857142857143	0.0974740575962738\\
32.8571428571428	0.0880222459544865\\
36.4285714285714	0.0880222459544864\\
40	0.0880222459544865\\
};

\addplot[fill = mycolor1, area legend] fill between[of=A and B];
    
\addplot [color=mycolor2, dashed, line width=\lw pt, mark size=\ms pt, mark=square, mark options={solid, mycolor2}]
  table[row sep=crcr]{%
-10	13.8237863084718\\
-6.42857142857144	9.15096923499022\\
-2.85714285714283	6.05373982181038\\
0.714285714285722	4.00103988359725\\
4.28571428571428	2.64090251709951\\
7.85714285714283	1.7400981931945\\
11.4285714285714	1.14413478244431\\
15	0.750750769255665\\
18.5714285714286	0.492359283759677\\
22.1428571428572	0.32446122025767\\
25.7142857142857	0.21818275089981\\
29.2857142857143	0.156393716476203\\
32.8571428571428	0.141228615252915\\
36.4285714285714	0.141228615252915\\
40	0.141228615252915\\
};

\addplot [name path = C, color=mycolor2, line width=\lw pt, forget plot]
  table[row sep=crcr]{%
-10	19.3263348299784\\
-6.42857142857144	12.7935061717403\\
-2.85714285714283	8.46342674569417\\
0.714285714285722	5.59365102534246\\
4.28571428571428	3.69211192649289\\
7.85714285714283	2.43274306823649\\
11.4285714285714	1.59955683650811\\
15	1.04958659058604\\
18.5714285714286	0.688342554076979\\
22.1428571428572	0.453612783221351\\
25.7142857142857	0.305030243084077\\
29.2857142857143	0.21864612650092\\
32.8571428571428	0.197444567287568\\
36.4285714285714	0.197444567287568\\
40	0.197444567287568\\
};
\addplot [name path = D, color=mycolor2, line width=\lw pt, forget plot]
  table[row sep=crcr]{%
-10	8.61582276571573\\
-6.42857142857144	5.70343951388156\\
-2.85714285714283	3.77305813404458\\
0.714285714285722	2.49369092854884\\
4.28571428571428	1.64597075802001\\
7.85714285714283	1.08453482229525\\
11.4285714285714	0.713094248251631\\
15	0.467913451842602\\
18.5714285714286	0.306868193074453\\
22.1428571428572	0.202223928069167\\
25.7142857142857	0.135984734597425\\
29.2857142857143	0.0974740575962738\\
32.8571428571428	0.0880222459544865\\
36.4285714285714	0.0880222459544865\\
40	0.0880222459544865\\
};

\addplot[fill = mycolor2, area legend] fill between[of= C and D];

\addplot [color=black, line width=\lw pt]
  table[row sep=crcr]{%
-10	13.5877888307652\\
-6.42857142857144	9.00694187847235\\
-2.85714285714283	5.97043441082164\\
0.714285714285722	3.95762374564909\\
4.28571428571428	2.62339130361036\\
7.85714285714283	1.738968222895\\
11.4285714285714	1.15271041574198\\
15	0.764097517749913\\
18.5714285714286	0.50649756318526\\
22.1428571428572	0.33574219985435\\
25.7142857142857	0.22255353817331\\
29.2857142857143	0.147524134216509\\
32.8571428571428	0.0977893694926681\\
36.4285714285714	0.0648216702749061\\
40	0.0429683610706142\\
};

\end{axis}

\begin{axis}[%
width=1.25\fwidth,
height=1.25\fheight,
at={(-0.163\fwidth,-0.156\fheight)},
scale only axis,
xmin=0,
xmax=1,
ymin=0,
ymax=1,
axis line style={draw=none},
ticks=none,
axis x line*=bottom,
axis y line*=left
]
\end{axis}
\end{tikzpicture}%
\vspace{-0.5cm}
\end{minipage}
\hfill
\begin{minipage}[b]{0.49\linewidth}
\centering
\setlength\fheight{2.3 cm} 
\setlength\fwidth{3 cm}
%
%
\definecolor{mycolor1}{rgb}{0.00000,1.00000,1.00000}%
\definecolor{mycolor2}{rgb}{1.00000,0.00000,1.00000}%
\newcommand\ms{0.8}
\newcommand\lw{0.3}

\pgfplotsset{every tick label/.append style={font=\tiny}}

\begin{tikzpicture}

\begin{axis}[%
width=0.969\fwidth,
height=\fheight,
at={(0\fwidth,0\fheight)},
scale only axis,
xmin=0,
xmax=40,
xlabel style={font=\tiny, yshift=1 ex},
xlabel={$P$ [dBm]},
ymode=log,
ymin=0.000316227766016838,
ymax=1,
yminorticks=true,
ylabel style={font=\tiny, yshift=-1.5 ex},
ylabel={DEB [m]},
axis background/.style={fill=white},
xmajorgrids,
ymajorgrids,
legend style={font=\tiny, at={(0.03,1.23)}, anchor=south west, legend cell align=left, align=left, draw=white!5!black}
]

\addplot [color=blue, line width=\lw pt, mark size=\ms pt, mark=o, mark options={solid, blue}]
  table[row sep=crcr]{%
-10	0.28544618178781\\
-6.42857142857144	0.18895758148293\\
-2.85714285714283	0.125003156089991\\
0.714285714285722	0.0826171305363467\\
4.28571428571428	0.0545317703238721\\
7.85714285714283	0.0359311388428244\\
11.4285714285714	0.023625141318859\\
15	0.0155021884580897\\
18.5714285714286	0.0101666847621082\\
22.1428571428572	0.00669977200125021\\
25.7142857142857	0.00450523697245985\\
29.2857142857143	0.00322936048254587\\
32.8571428571428	0.00291621760374135\\
36.4285714285714	0.00291621760374135\\
40	0.00291621760374135\\
};

\addplot [name path = A, color=mycolor1, line width=\lw pt, forget plot]
  table[row sep=crcr]{%
-10	0.386718012441188\\
-6.42857142857144	0.255996769300255\\
-2.85714285714283	0.169352316325366\\
0.714285714285722	0.111928393347228\\
4.28571428571428	0.0738787875965478\\
7.85714285714283	0.0486789436489104\\
11.4285714285714	0.0320069711118536\\
15	0.0210021219112247\\
18.5714285714286	0.013773665143089\\
22.1428571428572	0.00907674608188872\\
25.7142857142857	0.00610362442634234\\
29.2857142857143	0.00437508695840466\\
32.8571428571428	0.00395084589501774\\
36.4285714285714	0.00395084589501774\\
40	0.00395084589501774\\
};

\addplot [name path = B, color=mycolor1, line width=\lw pt, forget plot]
  table[row sep=crcr]{%
-10	0.226701050678952\\
-6.42857142857144	0.150069908056354\\
-2.85714285714283	0.0992773721697628\\
0.714285714285722	0.0656144362463139\\
4.28571428571428	0.0433090733614888\\
7.85714285714283	0.0285364718376747\\
11.4285714285714	0.0187630618349115\\
15	0.0123118214062661\\
18.5714285714286	0.00807437010737432\\
22.1428571428572	0.00532095172014563\\
25.7142857142857	0.00357805435973055\\
29.2857142857143	0.00256475462319707\\
32.8571428571428	0.0023160568855255\\
36.4285714285714	0.0023160568855255\\
40	0.0023160568855255\\
};

\addplot[fill = mycolor1, area legend] fill between[of=A and B];

\addplot [color=mycolor2, dashed, line width=\lw pt, mark size=\ms pt, mark=square, mark options={solid, mycolor2}]
table[row sep=crcr]{%
-10	0.28544618178781\\
-6.42857142857144	0.18895758148293\\
-2.85714285714283	0.125003156089991\\
0.714285714285722	0.0826171305363467\\
4.28571428571428	0.0545317703238721\\
7.85714285714283	0.0359311388428244\\
11.4285714285714	0.023625141318859\\
15	0.0155021884580897\\
18.5714285714286	0.0101666847621082\\
22.1428571428572	0.0066997720012502\\
25.7142857142857	0.00450523697245985\\
29.2857142857143	0.00322936048254587\\
32.8571428571428	0.00291621760374135\\
36.4285714285714	0.00291621760374135\\
40	0.00291621760374135\\
};

\addplot [name path = C, color=mycolor2, line width=\lw pt, forget plot]
table[row sep=crcr]{%
-10	0.386718012441189\\
-6.42857142857144	0.255996769300255\\
-2.85714285714283	0.169352316325366\\
0.714285714285722	0.111928393347227\\
4.28571428571428	0.0738787875965478\\
7.85714285714283	0.0486789436489104\\
11.4285714285714	0.0320069711118536\\
15	0.0210021219112247\\
18.5714285714286	0.013773665143089\\
22.1428571428572	0.00907674608188873\\
25.7142857142857	0.00610362442634233\\
29.2857142857143	0.00437508695840467\\
32.8571428571428	0.00395084589501775\\
36.4285714285714	0.00395084589501775\\
40	0.00395084589501775\\
};
\addplot [name path = D, color=mycolor2, line width=\lw pt, forget plot]
  table[row sep=crcr]{%
-10	0.226701050678953\\
-6.42857142857144	0.150069908056354\\
-2.85714285714283	0.0992773721697627\\
0.714285714285722	0.0656144362463139\\
4.28571428571428	0.0433090733614888\\
7.85714285714283	0.0285364718376747\\
11.4285714285714	0.0187630618349115\\
15	0.0123118214062661\\
18.5714285714286	0.00807437010737432\\
22.1428571428572	0.00532095172014563\\
25.7142857142857	0.00357805435973055\\
29.2857142857143	0.00256475462319707\\
32.8571428571428	0.0023160568855255\\
36.4285714285714	0.0023160568855255\\
40	0.0023160568855255\\
};
\addplot[fill = mycolor2, area legend] fill between[of= C and D];

\addplot [color=black, line width=\lw pt]
 table[row sep=crcr]{%
-10	0.274489227520607\\
-6.42857142857144	0.181950761035315\\
-2.85714285714283	0.120609758497151\\
0.714285714285722	0.0799486287497193\\
4.28571428571428	0.0529955728176954\\
7.85714285714283	0.0351291921099462\\
11.4285714285714	0.0232860986811608\\
15	0.0154356635954413\\
18.5714285714286	0.0102318432079991\\
22.1428571428572	0.00678238514241761\\
25.7142857142857	0.00449584178382684\\
29.2857142857143	0.00298016006475245\\
32.8571428571428	0.00197545964439735\\
36.4285714285714	0.00130947355908772\\
40	0.00086801115214529\\
};

\end{axis}

\begin{axis}[%
width=1.25\fwidth,
height=1.25\fheight,
at={(-0.163\fwidth,-0.156\fheight)},
scale only axis,
xmin=0,
xmax=1,
ymin=0,
ymax=1,
axis line style={draw=none},
ticks=none,
axis x line*=bottom,
axis y line*=left
]
\end{axis}
\end{tikzpicture}%
\vspace{-0.5cm}
\end{minipage}
\centerline{(d) PAN ($x_\text{clip}=\unit[25]{V}$)}\medskip
\hfill
\caption{AEB (left) and DEB (right) for multiple realizations: (a) Phase noise, (b) Carrier frequency offset, (c) Mutual coupling, (d) Power amplifier nonlinearity. 
}
\label{fig:single_impairment_on_localization}\vspace{-5mm}
\end{figure}
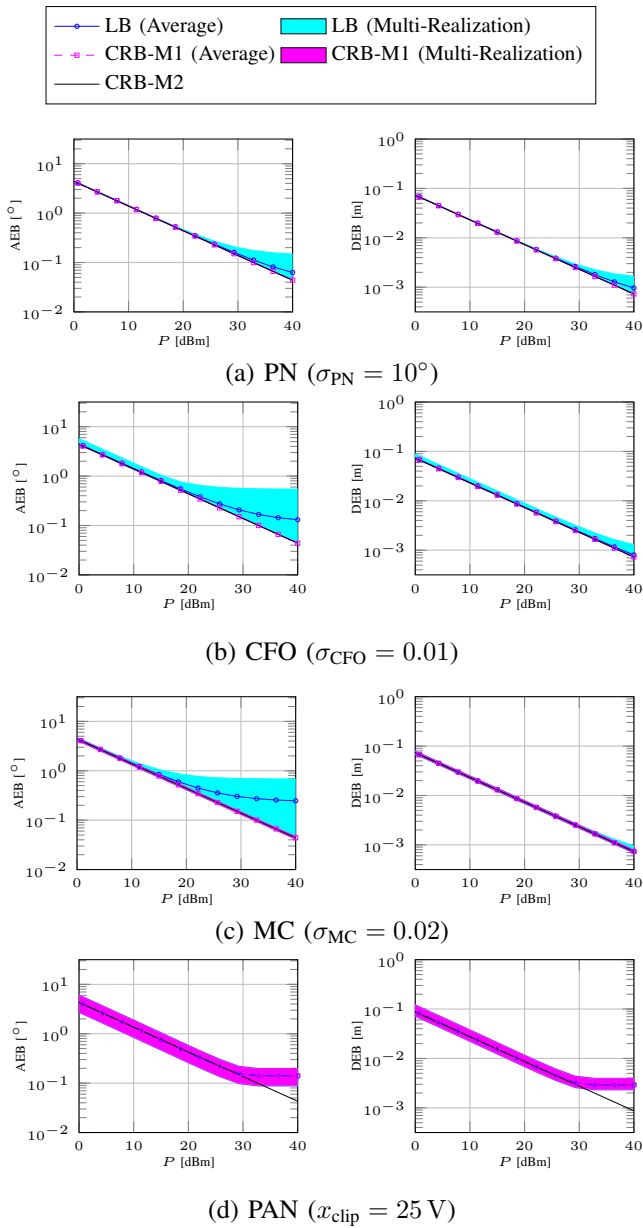
\vspace{-0.08cm}

\subsection{The Effect of Individual Impairments}
To gain a deeper understanding, we study the impact of \acp{hwi} on angle and delay estimation, considering  different \acp{hwi} separately. The results are shown in Fig.~\ref{fig:single_impairment_on_localization} for (a) PN, (b) CFO, (c) MC, and (d) PAN. 
Multiple realizations (multiple hardware realizations with a fixed pilot signal for (a)-(c) and multiple pilot signal realizations for (d)) are performed for each type of the HWI. We can see that different types of the \acp{hwi} affect angle and delay estimation differently. The PN and PAN affect both angle and delay estimation, however, the effect of PAN is mainly caused by the distortion of the pilot signals, whereas the effect of PN depends on the residual noise level $\sigma_\text{PN}$. The CFO and MC have a more significant effect on angle estimation compared to delay estimation since the CFO affects the phase changes across beams and the MC distorts the steering vector.
In addition, within reasonable levels of hardware impairments, the CRBs with perfect knowledge of the impairments (CRB-M1) are close to the CRBs of the \ac{mm} (CRB-M2).

\begin{figure}
\centering
%
%
\begin{tikzpicture}[scale=1\columnwidth/10cm,font=\footnotesize]
\begin{axis}[%
width=8cm,
height=4cm,
scale only axis,
xmin=0,
xmax=31,
xlabel style={font=\color{white!15!black}},
xlabel={$\sigma_\mathrm{PN} [^\circ]$},
ymode=log,
ymin=0.01,
ymax=0.199526231496888,
yminorticks=true,
ylabel style={font=\color{white!15!black}},
ylabel={PEB [m]},
axis background/.style={fill=white},
xmajorgrids,
ymajorgrids,
yminorgrids,
legend style={font=\footnotesize, at={(0.77,0.53)}, anchor=south east, legend cell align=left, align=left, draw=white!15!black, legend columns=2}
]
\addplot [color=red, dashed, line width=1.0pt]
  table[row sep=crcr]{%
0.1	0.132385379090722\\
4.37142857142857	0.132385379090722\\
8.64285714285714	0.132385379090722\\
12.9142857142857	0.132385379090722\\
17.1857142857143	0.132385379090722\\
21.4571428571429	0.132385379090722\\
25.7285714285714	0.132385379090722\\
30	0.132385379090722\\
};
\addlegendentry{CRB-M2 (10 dBm)}

\addplot [color=red, line width=1.0pt, mark=square, mark options={solid, red}]
 plot [error bars/.cd, y dir=both, y explicit, error bar style={line width=0.5pt}, error mark options={line width=0.5pt, mark size=2.0pt, rotate=90}]
 table[row sep=crcr, y error plus index=2, y error minus index=3]{%
0.097	0.132385578544556	4.07898129631268e-06	4.46549483268388e-06\\
4.36842857142857	0.132782352687293	0.000241929084937142	0.000179608303121553\\
8.63985714285714	0.13392516914591	0.000578550530538874	0.00049651477562318\\
12.9112857142857	0.135748266452907	0.00086506891431265	0.000939491728339376\\
17.1827142857143	0.13849491066994	0.00241119445321833	0.00141797133991894\\
21.4541428571429	0.142346523998492	0.00329701060872481	0.00298194260814627\\
25.7255714285714	0.146627426404942	0.00403840648112347	0.00332588968136405\\
29.997	0.151795394294114	0.00541085953902293	0.00388384252612164\\
};
\addlegendentry{LB (10 dBm)}

\addplot [color=blue, dashed, line width=1.0pt]
  table[row sep=crcr]{%
0.1	0.0418639326831513\\
4.37142857142857	0.0418639326831513\\
8.64285714285714	0.0418639326831513\\
12.9142857142857	0.0418639326831513\\
17.1857142857143	0.0418639326831513\\
21.4571428571429	0.0418639326831513\\
25.7285714285714	0.0418639326831513\\
30	0.0418639326831513\\
};
\addlegendentry{CRB-M2 (20 dBm)}

\addplot [color=blue, line width=1.0pt, mark=o, mark options={solid, blue}]
 plot [error bars/.cd, y dir=both, y explicit, error bar style={line width=0.5pt}, error mark options={line width=0.5pt, mark size=2.0pt, rotate=90}]
 table[row sep=crcr, y error plus index=2, y error minus index=3]{%
0.1	0.0418640078533722	1.30234903669008e-06	1.41933274018513e-06\\
4.37142857142857	0.042007651011311	0.000136234611851936	7.13745223520634e-05\\
8.64285714285714	0.0424206251174699	0.000807369510191468	0.000192346587161163\\
12.9142857142857	0.0431152575747449	0.0012185655561157	0.000451897384382872\\
17.1857142857143	0.0441574324280815	0.00254744860045827	0.000651876296406478\\
21.4571428571429	0.0456329930140885	0.00422712751844981	0.00147613670214539\\
25.7285714285714	0.047273469550474	0.00445178025127032	0.00162622059300851\\
30	0.048975391447078	0.00847166914252847	0.00205314345334793\\
};
\addlegendentry{LB (20 dBm)}

\addplot [color=black!30!green, dashed, line width=1.0pt]
  table[row sep=crcr]{%
0.1	0.0132385379090722\\
4.37142857142857	0.0132385379090722\\
8.64285714285714	0.0132385379090722\\
12.9142857142857	0.0132385379090722\\
17.1857142857143	0.0132385379090722\\
21.4571428571429	0.0132385379090722\\
25.7285714285714	0.0132385379090722\\
30	0.0132385379090722\\
};
\addlegendentry{CRB-M2 (30 dBm)}

\addplot [color=black!30!green, line width=1.0pt, mark=triangle, mark options={solid, black!30!green}]
 plot [error bars/.cd, y dir=both, y explicit, error bar style={line width=0.5pt}, error mark options={line width=0.5pt, mark size=2.0pt, rotate=90}]
 table[row sep=crcr, y error plus index=2, y error minus index=3]{%
0.103	0.0132385999350707	4.66901716785623e-07	4.71661626262609e-07\\
4.37442857142857	0.0133410533155805	0.000430924553464957	6.84494598662076e-05\\
8.64585714285714	0.0136296719358172	0.00210266228906429	0.000257700431722968\\
12.9172857142857	0.0141910512468272	0.00343782792220166	0.00066532243677307\\
17.1887142857143	0.0150097063757418	0.00575987332866312	0.00123836520900138\\
21.4601428571429	0.016122176488126	0.00995955853981913	0.00208633961260801\\
25.7315714285714	0.0173309118063106	0.00988128589483773	0.002826962228235\\
30.003	0.0180503428361922	0.0170614229375546	0.00311424291373025\\
};
\addlegendentry{LB (30 dBm)}

\end{axis}

\end{tikzpicture}%
\vspace*{-1 cm}
\caption{PEB vs. phase noise with different transmission power. A large PEB variance is observed in the scenarios with high SNR or large PN level.}
\label{fig:peb_vs_pn_noise}\vspace{-5mm}
\end{figure}
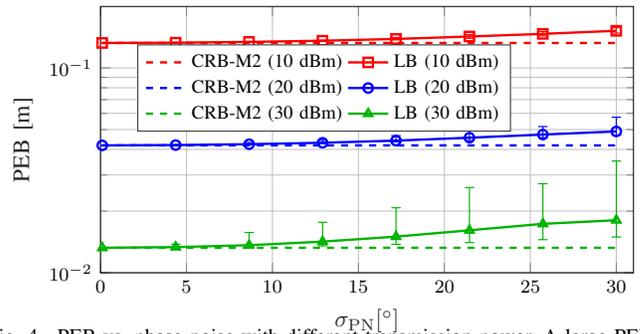

In Fig.~\ref{fig:peb_vs_pn_noise}, we  evaluate the localization performance (which in turn depends on the AEB (in terms of degree) and DEB (in terms of meter) shown in Fig.~\ref{fig:single_impairment_on_localization}) as a function of the standard deviation of PN $\sigma_\text{PN}$ (similar patterns can be seen for CFO and MC) for various values of average transmission power using 100 realizations of PN.
The impairments produce a larger perturbation in the high SNR scenario. As for low SNR, the noise level is high enough and the impairments, under a certain level, will not degrade the PEB too much.

\section{Conclusion}
\label{sec:conc}
\acp{hwi} present a crucial roadblock to achieving high performance in
radio-based localization. We modeled different types of \acp{hwi} and utilize the \ac{mcrb} to evaluate the error caused by model-mismatch. The effects of residual PN, residual CFO, residual MC and PAN on angle/delay estimation are evaluated. We found that PN and PAN affect both angle and delay estimation, whereas CFO and MC have a more significant effect on angle estimation.
We also observed that with perfect knowledge of the \acp{hwi}, the bound is close to the bound of the \ac{mm}, but will saturate at a certain level in the high SNR regime due to the PAN. In conclusion, dedicated pilot signal design, HWIs estimation and mitigation algorithms are needed for accurate localization in 6G systems.

\section*{Acknowledgment}
This work was supported, in part, by the European Commission through the H2020 project Hexa-X (Grant Agreement no. 101015956) and by the MSCA-IF grant 888913 (OTFS-RADCOM).

\appendices


\section{}
\label{appendix_B}
The pseudo-true parameters can be calculated as
\footnotesize
\begin{align*}
    {\boldsymbol\theta}_0 & = \arg\min_{\boldsymbol{\theta}} D_\text{KL}(f_\text{TM}(\yv|\bar {\boldsymbol\theta})\Vert f_\text{MM}(\yv, {\boldsymbol\theta})).
\end{align*}
\normalsize
After some manipulation, we find that 
\footnotesize
\begin{align*}
\footnotesize
    {\boldsymbol\theta}_0 
    & = \arg\min_{\boldsymbol{\theta}} \int {f_\text{TM}(\yv|\bar {\boldsymbol\theta})}\Vert \yv - \muv({\boldsymbol\theta})\Vert^2 \text{d} \yv\\
    & = \arg\min_{\boldsymbol{\theta}} \sum_{g,k} \int {f_\text{TM}(y_{g,k}|\bar {\boldsymbol\theta})}\vert y_{g,k} - \mu_{g, k}({\boldsymbol\theta})\vert^2 \text{d} y_{g,k}.
\end{align*}
\normalsize
For each received symbol at the $g$th transmission and $k$th subcarrier, $y_{g,k}\sim \mathcal{CN}(\bar\mu_{g,t}(\bar {\boldsymbol\theta}), \sigma_n^2)$, by ignoring the indices $g$ and $k$ we can have
\footnotesize
\begin{align*}
    \int & {f_\text{TM}(y|\bar {\boldsymbol\theta})}\vert y - \mu({\boldsymbol\theta})\vert^2 d y = \int {f_\text{TM}(y|\bar {\boldsymbol\theta})}\vert y - \bar\mu(\bar {\boldsymbol\theta})\vert^2 \text{d} y \\
    & + \int {f_\text{TM}(y|\bar {\boldsymbol\theta})}\vert \bar\mu(\bar {\boldsymbol\theta}) - \mu({\boldsymbol\theta})\vert^2 \text{d} y \\
    & + \int {f_\text{TM}(y|\bar {\boldsymbol\theta})}\vert (y - \bar\mu(\bar {\boldsymbol\theta}))(\bar\mu(\bar {\boldsymbol\theta}) - \mu({\boldsymbol\theta}))\vert \text{d} y\\
    & = \sigma_n^2 + \vert \bar\mu(\bar {\boldsymbol\theta}) - \mu({\boldsymbol\theta})\vert^2 + 0,
\end{align*}
\normalsize
from which \eqref{eq:pseudotrue_final} follows immediately. 

\section{}
\label{appendix_C}
To obtain matrices $\Am$ and $\Bm$, we need to calculate the derivative of the noise-free version of the \ac{mm} $\muv(\thetav)$ with respect to the channel parameters inside the vector $\thetav$.

\subsection{First-order Derivatives}
For the first-order derivative parts of the matrix $\Am$, note that $\alpha$ is dependent of $\rho$ and $\xi$, $\av$ is dependent of \ac{aoa} $\vartheta$, $\Dm$ is dependent of delay $\tau$. We can write $\frac{\partial \mu_{g,k}}{\partial \vartheta}  = \wv_{\text{}g}^\top {\alpha} \tilde\Cm_\text{}\dot{\av}_{\text{}\vartheta} D_k$,  $\frac{\partial \mu_{g,k}}{\partial \tau}  = \wv_{\text{}g}^\top {\alpha} \tilde\Cm_\text{}\av_\text{} \dot{D}_{{k,\tau}}$, $\frac{\partial \mu_{g,k}}{\partial \rho}  = \wv_{\text{}g}^\top \dot{\alpha}_\rho \tilde\Cm_\text{}\av_\text{} D_k$, $\frac{\partial \mu_{g,k}}{\partial \xi}  = \wv_{\text{}g}^\top \dot{\alpha}_\xi \tilde\Cm_\text{}\av_\text{} D_k$. 
where $\dot{D}_{k,\tau}  = -j2\pi k\Delta_f D_k$, $\dot{\alpha}_{\rho}  = e^{-j\xi}$, $\dot{\alpha}_{\xi}  = -j\alpha$  and $\dot{\av}_{\text{}\vartheta}  = [0, {j\pi\cos(\vartheta)}, \ldots, {j(N-1)\pi\cos(\vartheta)}]^\top\odot \av_{\text{}}.$ 

\subsection{Second-order Derivatives}
The second-order derivative parts of the matrix $\Bm$ in equation~\eqref{eq:matrix_B} can be obtained based on the dependence of the channel components ($\alpha, \av_{\text{}}, D$) and unknowns ($\vartheta, \tau, \rho, \xi$) with several examples as $\frac{\partial^2 \mu_{g,k}}{\partial \vartheta \partial \vartheta}  = \wv_{\text{}g}^\top {\alpha} \tilde\Cm_\text{}\ddot{\av}_{\text{}\vartheta} D_k$, 
$\frac{\partial^2 \mu_{g,k}}{\partial \theta \partial \tau}  = \wv_{\text{}g}^\top {\alpha} \tilde\Cm_\text{}\dot{\av}_{\text{}\theta} \dot{D}_{k, \tau}$, 
 $\frac{\partial^2 \mu_{g,k}}{\partial \tau \partial \tau} = \wv_{\text{}g}^\top {\alpha} \tilde\Cm_\text{}\av_\text{} \ddot{D}_{{k,\tau}}$, 
 $\frac{\partial^2 \mu_{g,k}}{\partial \rho \partial \rho}  = \wv_{\text{}g}^\top \ddot{\alpha}_\rho \tilde\Cm_\text{}\av_\text{} D_k$, and $\frac{\partial^2 \mu_{g,k}}{\partial \rho \partial \xi}  = \wv_{\text{}g}^\top \ddot{\alpha}_{\rho, \xi} \tilde\Cm_\text{}\av_\text{} D_k$.
where $\ddot{\av}_{\text{}\vartheta} = [0, {j\pi\cos(\theta)}, \ldots, {j(N-1)\pi\cos(\theta)}]^\top\odot \dot\av_{\text{}}$, $\ddot{D}_{k,\tau} = -j2\pi k\Delta_f \dot{D}_k$, $\ddot{\alpha}_\rho  = 0$, $\ddot{\alpha}_\xi  = -\alpha$, and $\ddot{\alpha}_{\rho, \xi}  = -j e^{-j\xi}$.




\ifCLASSOPTIONcaptionsoff
\newpage
\fi
\balance 
\bibliographystyle{IEEEtran}
\bibliography{IEEEabrv, ref}

\end{document}